\documentclass[letterpaper,11pt]{article}

\usepackage{citesort}
\usepackage{array}
\usepackage{color}
\usepackage{bm}
\usepackage{graphicx}

\evensidemargin \oddsidemargin

\newcommand{\prd}{\partial}

\newcommand{\ga}{\gamma}

\newcommand{\ba}{\begin{eqnarray}}
\newcommand{\ea}{\end{eqnarray}}
\newcommand{\be}{\begin{equation}}
\newcommand{\ee}{\end{equation}}
\newcommand{\begarrr}{\begin{array}{l}}
\newcommand{\ear}{\end{array}}
\newcommand{\om}{\omega}
\newcommand{\omc}{\Omega}

\newcommand{\Nk}{N_{\mybk}}
\newcommand{\nk}{n_{\mybk}}

\newcommand{\vb}{{\bf v}}
\newcommand{\ec}{{\cal E}}
\newcommand{\nc}{{\cal N}}

\newcommand{\mybk}{ {\bm{k}} }
\title{Introduction to Nonlinear Phenomena in Superfluid Liquids and
Bose-Einstein Condensates: Helium, Semiconductors and Graphene}

\author{Oleg L. Berman$^{a}$, Roman Ya. Kezerashvili$^{a,b}$, and German 
V. Kolmakov$^{c,d,e}$\thanks{Corresponding author. Email: gek11@pitt.edu} \\ \\
\small $^{a}$\sl{Physics Department, New York City College of Technology, City
University of New York, USA}\\
\small $^{b}$\sl{Graduate School and University Center, City University of New York, USA}\\
\small $^{c}$\sl{Department of Physics, University of Lancaster, UK}\\
\small $^{d}$\sl{Department of Chemistry, University of Pittsburgh, USA}\\
\small $^{e}$\sl{Center for Simulation and Modeling, University of Pittsburgh, USA}}

\begin{document}
\maketitle

\begin{abstract}
We review current understanding of the non-equilibrium dynamics of collective
quantum systems. We describe an approach based on the Hamiltonian formulation
of superfluid hydrodynamics. It is shown that, in the presence of constant
energy pumping, the nonlinear coupling of fluctuations in the density and
entropy strongly affects the nonequilibrium dynamics of the system. We use the
results obtained to analyze the properties of out-of-equilibrium superfluid
$^4$He and of exciton polariton Bose-Einstein condensates, both in
semiconductor quantum wells and in graphene layers in presence of high magnetic
field.
\end{abstract}\bigskip


\section{Introduction}\label{introduction}
Superfluidity is one of the most fascinating macroscopic quantum effects in
condensed matter. It exhibits itself as a frictionless, collective movement of
fluids -- either liquids or gases. After being discovered in 1938 by Kapitza
\cite{Kapitza:38} and by Allen and Misener \cite{Allen:38a} in their famous
experiments on helium $^4$He (the stable isotope of helium, of atomic mass
equal to 4) at temperatures below $\sim 2$ K, this phenomenon exerted an
influence on the development of condensed matter physics for the rest of the
20$^{\rm th}$ century. It is worthy of note that superconductivity (that is, a
complete loss of electrical resistivity at low temperatures) was later
explained as the superfluid motion of electronic Cooper pairs in metals \cite
{Landau:41,Schafroth:55}. We also mention that the apparent formation of a
superfluid fraction  in solids, called supersolid, has been recently observed
at mK temperatures \cite{Kim:04a}. The fundamentals of the theory of superfluid
liquids were developed in the pioneering works of Lev Landau \cite{Landau:41}
and Isaac Khalatnikov \cite{Khalatnikov:65}. It was subsequently established
that superfluidity can be understood in terms of Bose-Einstein condensation
(BEC) of particles and establishment of a long-range order in the system, known
as quantum coherence \cite{London:54,Feynman:72}. In particular, atoms in the
ground state of a supefluid are described by a single macroscopic wave function
$\Psi = \sqrt{\rho} \exp (i \theta)$, where $\rho$ is the condensate density
and $\theta$ is the phase.

In the last few years, interest in superfluid systems has further increased
because of the discovery of Bose-Einstein condensation in atomic gases at
$\mu$K temperatures \cite{Davis:95,Anderson:95}. For the discovery and study of
this phenomenon Cornell, Ketterle and Wieman shared the 2001 Nobel Prize in
Physics. Furthermore, it was recently found that excitons, which are bound
electron-hole pairs in semiconductor nanostructures, can also form a state
where macroscopic quantum coherence strongly affects their dynamics
\cite{Berman:04,Berman:08,Berman:08b}. Other examples of systems in which
quantum coherence is important are provided by polaritons, which are a resonant
coupling of excitons with cavity photons in semiconductors
\cite{Kasprzak:06,Amo:09,Berman:08a}, and by magnitopolaritons, which are
generated in graphene embedded in a microcavity when magnitoexcitons are
coupled with photons in a strong magnetic field
\cite{Berman_magnetopolaritons}.  Geim and Novoselov have received the 2010 
Nobel Prize in Physics for the discovery and study of graphene.

It was found that, under certain conditions,
the polaritons form a superfluid liquid that is in many aspects similar to
superfluid $^4$He \cite{Berman:08a}. Because of the extremely small polariton effective mass, the
latter can exist as a superfluid at temperatures much higher than that of the
temperature of normal-to-superfluid transition in liquid $^4$He. 
The example of BEC and superfluidity of magnitoexcitons in
graphene in strong magnetic fields is of special importance because graphene is
considered as a hot candidate for future applications in electronics.

The key feature of the exciton polariton superfluid in microcavities is that it
can {\it only} form under nonequilibrium conditions: the composite particles
have a finite lifetime and the particle losses must be compensated by
continuous production (typically, resonance pumping by laser radiation is used
\cite{Carusotto:04}). Because of that, there is currently a strong interest in
kinetic phenomena in non-equilibrium superfluid liquids
\cite{Griffin:09,Zakharov:05,YLvov:03}. Although out-of-equilibrium superfluids
can be studied, in principle, in a number of systems, superfluid helium $^4$He
in a macroscopic cavity presents a useful testbed for experimental studies of
general kinetic and transport properties. This potentially fruitful approach
has already yielded important results \cite{Kolmakov:06,Ganshin:08,Efimov:10}.
Specifically, it has been demonstrated that the main mechanisms of energy
relaxation in the superfluid liquid is governed by the nonlinear dynamics of
second sound oscillations and its coupling with ordinary sound, i.e.\ with
density waves in the bulk superfluid. (Second sound in a quantum system is a
slightly dissipative entropy wave that can be understood as a density wave in
the gas of quasiparticles \cite{Khalatnikov:65}: see Sec. \ref{sec:helium}).
Non-equilibrium conditions were achieved in these experiments through
continuous driving at a resonant frequency of the cavity by a periodic drive
applied to a heater immersed in the superfluid \cite{Kolmakov:06}. It was found
that, because of strong nonlinearity of the second sound interactions, the
perturbations play an important role in the non-equilibrium dynamics of the
system; this interaction results in the development of energy cascades and the
formation of constant fluxes of conserved quantities (e.g.\ energy and wave
action) through the frequency scales (cf.\ Sec. \ref{sec:kinetics}).

In this paper, we review recent achievements in the theory of non-equilibrium
superfluid systems. In Sec. \ref{sec:helium}, we develop a general formalism
that describes its nonlinear dynamics. Through this technique, we find
steady-state non-equilibrium solutions of the proper kinetic equations
describing the correlation functions for the density and entropy fluctuations.
Here, we take advantage of the Hamiltomian approach to describe of nonlinear
dynamics of non-equilibrium superfluid liquids. In particular, we utilize an
approach based on the kinetic equation for the correlation functions for the
density and entropy fluctuations which is a generalization of that used in
consideration of classical nonlinear turbulent systems. Through analytical
computations and numerical simulations, we describe the concept of second sound
turbulence, which was observed in recent experiments \cite{Kolmakov:06}. In
Sec. \ref{sec:polaritons}, we focus on application of the Hamiltonian formalism
to a particular system of exciton polariton supefluids in microcavities. In
doing so, we demonstrate the transition to a superfluid state in this system
and describe the main characteristics of the Bose-Einstein condensate fraction
as functions of the interaction parameters of exciton polaritons with each
other and with an external trapping field.

\section{Nonlinear phenomena in superfluid liquid}\label{sec:helium}
\subsection{Hamiltonian formulation of superfluid hydrodynamics in quantum liquids}\label{superfluidhydro}
\subsubsection{Superfluid hydrodynamics}
As mentioned above in the Introduction, the Hamiltonian formalism plays a
central role in the subsequent consideration of kinetic processes in quantum
superfluid systems. For the reader's convenience, we now sketch the Hamiltonian
formulation of the superfluid hydrodynamics equations
\cite{Pokrovskii:76,Khalatnikov:95,Kolmakov:03,Kolmakov:04a,Kolmakov:06,Kolmakov:09}. 
The Landau's well-known hydrodynamics equations for a quantum
liquid that has the total density $\rho$ and the densities $\rho_s$ and
$\rho_n$  for its superfluid and normal components are \cite{Khalatnikov:65}
\begin{equation}\setlength{\extrarowheight}{5.mm}
\begin{array}{ll} \displaystyle
 \frac{\partial \rho }{\partial t } + \mbox{\rm div } \,\mbox{\bf j} = 0, &  \displaystyle
 \quad  \frac{\partial S }{\partial t } + \mbox{\rm div} \,(S \bm{v}_{n}) = 0,  \\
 \displaystyle  \frac{\partial \bm{v}_{s} }{\partial t} +
   \nabla \left( \mu + \frac{1}{2}\, \bm{v}_{s} ^{2} \right) = 0, & \displaystyle \quad
 \frac{\partial \mbox{\it j}_{\mbox{\it i} } }{\partial t} +
   \frac{\partial \Pi _{\mbox{\it ik}}}{\partial  x _{\mbox{\it k}}} = 0.
\end{array} \label{superflueq}\setlength{\extrarowheight}{0.mm}
\end{equation}
Here $S$ is the entropy per unit volume, $\bm{v}_{s}$ and $\bm{v}_{n}$ are the
velocities of superfluid and normal motion, $\bm{j} =\rho _{s} \bm{v}_{s} +
\rho _{n} \bm{v}_{n}$ is the mass flux, $\Pi _{{ik}} = p\,\delta _{{ik}} + \rho
_{s}v_{si}v_{s k} + \rho _{n}\,v_{ni}\,v_{nk}$ is the tensor of the momentum
flux, $\mu = \partial \varepsilon / \partial \rho$ is the chemical potential of
the unit mass, and $\varepsilon$ is the energy per unit volume.  Equations
(\ref{superflueq}) are the continuity equations for the density, entropy and
momentum of the fluid, and the equation for the superfluid velocity, which is
the phase gradient of the condensate wave function, $\bm{v}_s = \hbar \nabla
\theta /m_4$, where $\hbar$ is Planck's constant and $m_4$ is the mass of a
$^4$He atom. The damping terms in Eqs.\ (\ref{superflueq}) are omitted. In this
review, we restrict our consideration to processes for which the thermal
expansion of the fluid is not an essential feature, and the terms proportional
to the thermal expansion coefficient $\kappa_T = (T/\rho) (\partial \rho /
\partial T)$ of the fluid are therefore neglected. For superfluid $^4$He, this
coefficient is indeed small, with $\kappa_T \sim 10^{-2}$ over the range of
temperature $T<2.15$~K.

In a non-stationary state, the density, entropy and velocities of the
superfluid are not equal to their equilibrium values. For an infinitely small
deviation from equilibrium, Eqs.\ (\ref{superflueq}) are reduced to two linear
wave equations for perturbations of the density $\delta \rho$, and the entropy
per unit mass $\delta \sigma$  \cite{Khalatnikov:65},
\begin{equation}
 {\partial^2 \delta \rho \over \partial t^2} = u_{1}^{2} \Delta
\delta \rho,  \qquad
 {\partial^2 \delta \sigma \over \partial t^2}= u_{2}^{2} \Delta
  \delta \sigma.  \label{lineareqs}
\end{equation}
Here $u_{1} = (\partial p / \partial \rho)^{1/2}$ and $u_2 = (\rho_{s} \sigma^2
T/ \rho_n c)^{1/2}$ are the first and second sound velocities, $\sigma =
S/\rho$, and $c$ is the heat capacity per unit mass. The linear equations
(\ref{lineareqs}) describe the propagation of two types of waves in the bulk
superfluid: density waves in which $\delta \sigma = 0$ at velocity $u_{1}$, and
entropy waves in which  $\delta \rho=0$ at velocity $u_2$. These waves are
known as {\it first sound} and {\it second sound},respectively
\cite{Khalatnikov:65}. In second sound, the fluid density remains constant but
the temperature varies with coordinates and time as $\delta T = (\partial T/
\partial \sigma) \delta \sigma$ \cite{Lifshitz:44}. The possibility of
propagating of temperature waves that are only weakly dissipative is a hallmark
of quantum systems, and is in sharp contrast to diffusive heat transfer
mechanisms in classical media where the temperature waves is damped over a
distance of the order of its own wavelength \cite{Landau:87}.

\subsubsection{Hamiltonian representation of  hydrodynamic
equations for the superfluid liquid in a cavity}\label{sec:represent} As we
have seen above, the dynamics of a superfluid liquid can be described, to a
first approximation, as the propagation of the density and entropy waves. In
what follows, we disregard modes that are localized at the fluid boundaries.
(For a detailed consideration of nonlinear phenomena at the surface of quantum
liquids, see Refs. \cite{Khalatnikov:95,Kolmakov:09,Abdurakhimov:09}). At
finite wave amplitude, however, nonlinear effects in the wave interaction play
an important role. In particular, second sound is characterized by rather
strong nonlinear properties \cite{Dessler:56,Osborne:50}. This can lead to the
formation of shock waves (temperature discontinuities) during the propagation
of finite amplitude waves at short distances from the source
\cite{Iznankin:83,Borisenko:88,MezhovDeglin:80,Goldner:90,Kolmakov:06b}. The
velocity of a traveling second sound wave depends on its amplitude and, to a
leading order approximation, can be written as $u_2 = u_{20} (1+ \delta T)$,
where $\delta T$ is the wave amplitude, $u_{20}$ is the velocity of a wave of
an infinitely small amplitude, and
\begin{equation}
 \alpha_2 = {\partial \over \partial T} \ln \left( u_{20}^3
 {\partial S \over \partial T} \right) \label{alphacoeff}
\end{equation}
is the nonlinearity coefficient. In He-II, the superfluid phase of liquid
$^4$He, the nonlinearity coefficient $\alpha_2$ of second sound may be either
positive and negative, depending on the temperature and pressure
\cite{Khalatnikov:65,Dessler:56,Efimov:99,Kolmakov:06b}; under the saturated
vapor pressure, the nonlinearity coefficient is positive ($\alpha_2>0$) at
temperatures $T<T_{\alpha} = 1.88$ K (like the nonlinearity of conventional
sound waves in ordinary media), but it is negative in the range
 $T_{\alpha} < T < T_{\lambda}$. Here, $T_{\lambda} = 2.176$ K
is the temperature of the superfluid-to-normal transition in bulk $^4$He.

To describe nonlinear effects, we now introduce the Hamiltonian variables for
the system (\ref{superflueq}). An arbitrary flow of superfluid can be described
by three pairs of conjugate variables $(\alpha,\rho)$, $(\beta,S)$ and
$(\gamma,f)$ \cite{Pokrovskii:76,Khalatnikov:95}. Here $\alpha = \hbar
\theta/m$ is the superfluid velocity potential, $\beta$ is the phase variable
conjugated to $S$, and $\gamma$ and $f$ are Clebsch variables. The Clebsch
variables define the vorticity in the system; we focus below on the potential
motion of the superfluid, so we set  $\gamma = f = 0$. The Hamiltonian function
of the superfluid system is given by the total energy expressed in the
stationary frame of reference,
\begin{equation}
 H= \int d^3 \bm{r} \left[ \frac{\rho _{n}}{2}\,\bm{v}_{n}^{2}  +
\frac{\rho _{s}}{2}\,\bm{v}_{s}^{2} +
\varepsilon (\rho, S, \bm{p}) \right]. \label{hamalpha}
\end{equation}
Here $\varepsilon (\rho, S, \bm{p})$ is the energy per unit volume of the
superfluid component in the reference frame moving with the velocity
$\bm{v}_s$, and $\bm{p}$ is the momentum of relative motion of the normal
component. The total mass flux is expressed through these variables as
$\mbox{\bf j} = \rho \mbox{\bf v} _{s} + \bm{p}  = \rho \nabla \alpha  + S
\nabla \beta$. The equations of superfluid hydrodynamics (\ref{superflueq}) for
the Hamiltonian function (\ref{hamalpha}) are derived in terms of the conjugate
Hamiltonian variables as follows \cite{Pokrovskii:76,Khalatnikov:95}
\begin{equation}
 \dot {\rho} = {\delta H \over \delta \alpha}, \qquad
 \dot {\alpha} = - {\delta H \over \delta \rho}, \qquad
 \dot {S} = {\delta H \over \delta \beta}, \qquad
 \dot {\beta} = - {\delta H \over \delta S}. \label{shorthameq}
\end{equation}
In Eqs.\ (\ref{shorthameq}) $\delta$  denotes the variational derivative
\cite{Ramon:81}. In this representation, the momentum of the relative motion of
the normal component is equal to $\bm{p} = S \nabla \beta$.

If the sound waves propagate in an unrestricted superfluid system, the
conjugate variables can be expressed in terms of normal coordinates -- the
normalized amplitudes of the running first and second sound waves
\cite{Pokrovskii:76}. In a cavity, the normal modes satisfy appropriate
boundary conditions. In the important case of a high-quality resonator
\cite{Borisenko:88,Kolmakov:06,Ganshin:08,Efimov:08b,Ganshin:10}, the
corresponding normal modes  are standing waves of first and second sound whose
frequencies are equal to the resonant frequencies, $\Omega_k = u_1
|{\bm{k}}_n|$, $\omega_k = u_2 |\bm{k}_n|$, where $\bm{k}_n$ is the $n$th
resonant wave vector.

Variations of the Hamiltonian variables in a wave with respect to their
equilibrium values are expressed via the sound amplitudes as
\cite{Pokrovskii:76,Khalatnikov:95,Brazhnikov:04a}

\begin{equation} \setlength{\extrarowheight}{2.mm}
\begin{array}{ll} \displaystyle
 \alpha(\bm{r},t) = \sum_{\bm{k}} \bar{\alpha} \varphi_{\bm{k}}(\bm{r}) (a_{\bm{k}}- a_{\bm{k}}^*), & \quad \displaystyle
 \delta \rho(\bm{r},t) = \sum_{\bm{k}} \bar{\rho} \varphi_{\bm{k}}(\bm{r})
(a_{\bm{k}} + a_{\bm{k}}^*), \\ \displaystyle
 \beta(\bm{r},t) = \sum_{\bm{k}} \bar{\beta} \varphi_{\bm{k}}(\bm{r}) (b_{\bm{k}}- b_{\bm{k}}^*), & \quad \displaystyle
 \delta S(\bm{r},t) = \sum_{\bm{k}} \bar{S} \varphi_{\bm{k}}(\bm{r}) (b_{\bm{k}}
+ b_{\bm{k}}^*),\label{eq:abrs}
\end{array}\setlength{\extrarowheight}{0.mm}
\end{equation}
where the normal coordinates $a_{\bm{k}}$ and $b_{\bm{k}}$  are the complex
amplitudes of the first and second sound waves, respectively, $\bar{\rho}$,
$\bar{S}$, $\bar{\alpha}$ and $\bar{\beta}$ are the normalizing factors, and a
star stands for the complex conjugate. The complex amplitudes are utilized to
take the wave phases into account.  The spatial basic functions depend on the
geometry of the cavity and the boundary conditions. For the simplest case of a
high quality rectangular resonator, they are $\varphi_{\bm{k}}(\bm{r}) =
\cos(k_x x) \cos(k_y y) \cos(k_z z)$.
The equations of motion (\ref{superflueq}) are presented by the Hamiltonian
equations for the sound amplitudes as follows:
\begin{equation}
 i \dot{a}_{\bm{k}} = {\partial H \over \partial a_{\bm{k}}^*} -i \gamma_{\bm{k}}^{(1)} a_{\bm{k}}
  + F _{\bm{k}}^{(1)}, \qquad
 i \dot{b}_{\bm{k}} = {\partial H \over \partial b_{\bm{k}}^*} -i \gamma_{\bm{k}}^{(2)} b_{\bm{k}}
  + F _{\bm{k}}^{(2)}. \label{abeqs}
\end{equation}
The dissipation $\gamma_{\bm{k}}^{(i)}$, and the interaction with an external
driving force $F _{\bm{k}}^{(i)}$, are included phenomenologically
\cite{Pokrovskii:76,Khalatnikov:95}; the superscript $i=1,2$ labels the first
and second sound modes, respectively. Note that this representation provides a
quasiclassical limit for the equations of motion of a superfluid system that
can be used provided that the occupation numbers of the corresponding states
are sufficiently large, i.e.\ $|a_{\bm{k}}|^2$, $|b_{\bm{k}}|^2 \gg \hbar$.
However, for the purely quantum case $|a_{\bm{k}}|^2$, $|b_{\bm{k}}|^2 \sim
\hbar$, the normal variables must be considered as operators in a proper
Hilbert space; we will consider examples of such systems in Sec.\
\ref{sec:polaritons}.

The nonlinear dynamics of the system (\ref{abeqs}) can be considered in terms
of an expansion of the Hamiltonian $H$ in a Taylor series over the wave
amplitudes $a_{\bm{k}}$ and $b_{\bm{k}}$, $H = H_2 + H_3 + \ldots$. The term
$H_2$ is a quadratic function of the amplitudes, $H_2 = \sum_{\bm{k}} \left(
\Omega_{\bm{k}} |a_{\bm{k}}|^2 + \omega_{\bm{k}} |b_{\bm{k}}|^2 \right)$, that
describes the propagation of linear waves \cite{Zakharov:92}. Nonlinear effects
such as parametric generation and decay of waves correspond to anharmonic terms
in the Hamiltonian, with the coefficients in the expansion over $a_{\bm{k}}$
and $b_{\bm{k}}$ being the amplitudes of the processes. Three-wave processes
include (\textit{a}) the decay of the first sound into two second sound waves,
(\textit{b}) Cerenkov emission of the second sound wave by the first sound
wave, and (\textit{c}) inner decay and confluence of waves that belong to the
same wave mode. As we show in Secs.\ \ref{sec:kinetics} and \ref{sec:numerics},
these processes are responsible for the generation of the direct and inverse
energy cascades in frequency space for a superfluid liquid.  In actual
experimental regimes, one of these processes may sometimes dominate (see Refs.\
\cite{Kolmakov:06,Ganshin:08,Nemirovsky:90,Rinberg:96,Rinberg:2001,Davidowitz:95}
and discussions in Sec.\ \ref{sec:numerics} below). In what
follows, we disregard interactions which involve four or more waves. Therefore,
the higher order terms in the expansion of the Hamiltonian over the wave
amplitudes are omitted. These higher-order processes are responsible for the
isotropisation in the system \cite{Landau:49,Khalatnikov:50} and should be
taken into account in a more general theory.

\subsubsection{Statistical description}\label{sec:ke}
Under experimental conditions
\cite{Brazhnikov:04a,Efimov:06,Efimov:07,Kolmakov:06,Ganshin:08,Ganshin:08b,Efimov:08,Efimov:08b}, 
many degrees of freedom (that is,
wave modes) in a superfluid liquid are excited simultaneously and interact with
each other. In this far-from-equilibrium state of a nonlinear system, the
dynamics should be described statistically. The theory of wave turbulence,
initially formulated for classical wave fields \cite{Zakharov:84,Zakharov:92},
presents a useful framework for this description. Below, we extend this
approach to the case of nonlinear \textit{quantum} systems and, in particular,
we study kinetic phenomena in a superfluid system described by the three-wave
Hamiltonian equations (\ref{hamalpha})--(\ref{abeqs}). In this statistical
approach, the state of the quantum system is characterized by simultaneous
correlation functions of the normal variables, averaged over the ensemble of
waves
\begin{equation}
 \langle a_{\bm{k}}^*(t) a_{\bm{k}_1}(t) \rangle = N_{\bm{k}}(t) \delta (\bm{k} -\bm{k}_1 ), \qquad
 \langle b_{\bm{k}}^*(t) b_{\bm{k}_1}(t) \rangle = n_{\bm{k}}(t) \delta (\bm{k} -\bm{k}_1 ). \label{nndef}
\end{equation}
The correlation functions $N_{\bm{k}}(t)$ and $n_{\bm{k}}(t)$ are the
``occupation numbers'' of the first and second sound wave modes. The time
evolution of the correlation functions obeys kinetic equations that are similar
to the Boltzmann equation \cite{Landau:statphys},
\begin{equation}
 \frac{\prd \Nk }{\prd t} = I_1 [\Nk, \nk], \qquad
 \frac{\prd \nk }{\prd t} = I_2 [\Nk, \nk], \label{ke}
\end{equation}
with $ I_1 [\Nk, \nk]$ and $I_2 [\Nk, \nk]$ being the ``collision''
integrals \cite{Zakharov:92,Nemirovsky:90,Pokrovskii:91,Kolmakov:95a}.

\subsection{Kinetic and nonstationary phenomena in superfluids}\label{sec:kinetics}
\subsubsection{Equilibrium fluctuations} \label{sec:eqf}
The kinetic Eqs.\ (\ref{ke}) can be used for studying the energy relaxation
processes in a superfluid. In Sec.\ \ref{sec:kinetics}, we focus on the case
where one of the parametric processes -- the decay of first sound to two second
sound waves and the inverse process of confluence of two second sound waves
into a first sound wave -- dominates. For superfluid $^4$He there is a large
window of bath temperatures and wave amplitudes, in which this process is
relevant \cite{Pokrovskii:76,Rinberg:2001}. The opposite situation where the
mutual transformation of the first and second sound waves is negligible, and
the wave dynamics is governed by the inner nonlinearity of the wave modes, is
described in Sec.\ \ref{sec:numerics}.

In the absence of an external energy input, the system tends towards
thermodynamic equilibrium where only temperature fluctuations exist. Such an
equilibrium state is described by a general Rayleigh-Jeans distribution
\cite{Kolmakov:95}
\begin{equation}
 \Nk = \frac{T}{ \omc_\mybk \, +\, (\vb_0 \mybk) \,+\, \mu}\, , \qquad
 \nk = \frac{T}{ \om_\mybk \, +\, (\vb_0 \mybk) \,+\, \mu^{\prime}}\, ,
 \label{td}
\end{equation}
which are the stationary solutions of the kinetic equations (\ref{ke}). The
vector $\vb_0$ is the constant drift velocity of the excitations in the
superfluid with respect to the stationary frame of reference, and $T$ is the
temperature. The term $ (\vb_0 \mybk)$ is simply the Galilean transformation to
a moving frame of reference. In Eqs.\ (\ref{td}) $\mu$, $\mu ^{\prime}$ play a
role of the effective chemical potentials for the quasiparticles, which obey
the relation $\mu = 2 \mu^{\prime}$. Let us note that $\mu$ and $\mu^{\prime}$
do not coincide with the chemical potential of atoms in the medium; the latter
always tends to zero for a Bose-Einstein condensed macroscopic superfluid
system. For the fluid at rest, one has $ \vb_0 = 0$. In this case, in the
low-frequency limit, $\omc_\mybk, \, \om_\mybk \, \ll \, \mu, \, \mu^{\prime}$,
one finds from Eqs.\ (\ref{td}) that $\Nk = \nk/2 =  T/\mu$. Thus, in
equilibrium, the long-wave distributions do not depend on $k$ and are
proportional to the temperature $T$ of the system.  In the high frequency
limit,  $\omc_\mybk, \, \om_\mybk \, \gg \, \mu, \, \mu^{\prime}$, one obtains
$\Nk, \, \nk \, \propto \, k^{-1}$.

\subsubsection{Direct and inverse cascades} \label{sec:cascades}
Next we turn our attention to the energy transfer mechanisms in a
non-equilibrium superfluid in presence of a continuous energy input. In
experiments with superfluid helium, this corresponds to the case where the
system is continuously driven by an external periodic source of heat or
pressure \cite{Kolmakov:06,Ganshin:08,Rinberg:2001,Davidowitz:95}. The energy
is absorbed in the superfluid through bulk dissipation and viscous drag at the
resonator walls. For the polariton Bose-Einstein condensates  the role of the
energy pump is played by laser radiation. In this case, the effective damping
is given by a finite lifetime of the photons in an optical cavity, see Sec.\
\ref{sec:polaritons} for a discussion. In both examples, the external driving
force is periodic, that is, localized in frequency space (we denote the latter
as $\omega_{\rm drive}$).

The steady, non-equilibrium state of the system is described by the equations
\begin{equation}
 I_1 [\Nk, \nk] = 0, \qquad I_2 [\Nk, \nk]=0. \label{ke0}
\end{equation}
An important property of Eq.\ (\ref{ke0}) is that the integrals $I_1$ and $I_2$
contain only power-law functions \cite{Kolmakov:95a}. Because of that, the
steady-state solutions of Eqs.\ (\ref{ke0}) acquire a power-law-like form \be
	\Nk\,=\,Ak^s, \qquad  \nk\,=\,Bk^s. \label{dist} \ee The scaling indices in
Eqs.\ (\ref{dist}) are equal to $s=-9/2$ and $s=-4$ \cite{Kolmakov:95a}. These
solutions describe the propagation of constant fluxes of two different
conserved quantities, or  the ``integrals of motion'' of Eqs.\ (\ref{ke}),
through the  scales. It can be shown that these are the wave energy, $\ec \,=\,
\int d\mybk \,[\omc_\mybk \Nk \,+\, \om_\mybk \nk]$, and the wave action $\nc
\,=\, \int d\mybk \,[2\Nk \,+\, \nk]$ (the last is sometimes also referred to
as ``the number of waves'' \cite{Zakharov:92}). To show that, we can rewrite
the kinetic equations (\ref{ke}) in the form of the continuity equations
\cite{Kolmakov:95a}
\begin{equation}
 \frac{\prd \ec}{\prd t} + \frac{\prd P}{\prd k}  = 0, \qquad
  \frac{\prd \nc}{\prd t} + \frac{\prd Q}{\prd k} = 0. \label{conteqs}
\end{equation}
From here it follows that both $\ec$ and $\nc$ are conserved over time.

The fluxes $P$ and $Q$ describe the propagation of the corresponding integrals
of motion in $K$-space. In the steady state, where $\partial \ec/ \partial t =
\partial \nc/ \partial t = 0$, the fluxes $P$ and $Q$ are constant
in accordance with Eqs.\ (\ref{conteqs}). For the non-equilibrium distributions
(\ref{dist}),  the energy flux is equal to $P = v_1 c_1 \rho^{-1} AB
\gamma_2^{11/2}$ at $s=-9/2$, and the flux of wave action is equal to $Q = -
v_2 \rho^{-1} \gamma_2 ^5 AB$ at  $s=-4$, where $\gamma_2 = u_2/u_1$ is the
second-to-first sound velocity ratio, and $v_1$ and $v_2$ are positive
dimensionless constants which depend on temperature \cite{Kolmakov:95a}. Thus,
in a steady state, the fluxes $P$ and $Q$ have opposite signs. Specifically,
the energy flux $P$ is positive and therefore, the corresponding integral of
motion, i.e. the wave energy $\ec$, propagates from the driving frequency scale
$\omega_{\rm drive}$ towards the high frequency spectral domain, i.e.\ the
\textit{direct cascade}. As a consequence, the steady-state distribution
(\ref{dist}) at $s=-9/2$ forms at frequencies higher than $\omega_{\rm drive}$.
The $Q$ flux is negative, so the wave action, $\nc$, is transferred towards the
low-frequency domain and results in formation of the \textit{inverse cascade}.
The corresponding solution (\ref{dist}) at $s=-4$ is established at frequencies
$\omega < \omega_{\rm drive}$.

Note that solutions of the type (\ref{dist}) are similar to the classical
turbulent spectra of surface waves on deep water \cite{Zakharov:92} or spectra
in optical turbulence \cite{Dyachenko:92}. In the latter cases, however, the
inverse cascade appears as a result of the direct four-wave interaction of
waves that explicitly preserves the number of waves \cite{Zakharov:92}. In a
quantum superfluid system, the interaction between sound waves is of the
three-wave type, and the corresponding integral, $\nc$, is conserved as the
result of nonlinear interaction between two wave modes -- the density and
entropy waves. Both the direct and inverse cascades have been observed in
recent experiments with superfluid $^4$He in a cavity
\cite{Ganshin:08,Efimov:08,Efimov:10}, see Sec.\ \ref{sec:exp}.

\subsubsection{Propagation of perturbations in $K$-space}\label{sec:diffapp}
We now consider the nonstationary processes through which the cascades are
formed. In doing so, we use the smallness of the parameter $\ga_2 = u_2/u_1$ to
simplify the description of the superfluid dynamics. Note that the value of
$\ga_2$ is typically small; for example, in He-II one has $\ga_2 \sim 0.1$ over
a wide range of temperatures \cite{Khalatnikov:65}.

It follows from the conservation of momentum ($\bm{k}$-vector) and energy
(frequency) in the decay process that, at small $\ga_2$, first sound of wave
vector $k$ only interacts with the second sound whose wave vector is close to
$k' = k/2 \ga_2$. This results in locality of the wave interaction in
$K$-space. As a consequence, the set of kinetic equations (\ref{ke}) can be
reduced to a Fokker-Plank differential equation that simplifies the analysis of
nonstationary processes in the superfluid liquid (cf.\ Eq.\ (\ref{dif})).

Another peculiarity of the kinetic equations  for the decay process is that the
characteristic evolution time for the first sound is very short compared to
that for second sound, $\tau_1/\tau_2 \sim \ga_2 ^3$. Because of this great
difference in time scales, we can simplify the whole system (\ref{ke}) by
considering the first sound waves as a ``fast'' subsystem, which is always in
quasi-equilibrium with the ``slow'' second sound subsystem. Consequently, at
large time, the wave dynamics is governed by the effective kinetic equation
$\dot{n}_\mybk = I_2[\Nk^{(qe)}, \nk]$, where $\Nk^{(qe)} = n
_{\bm{k^{\prime}}} /2$ is the quasi-equilibrium first sound distribution.
Taking into account the above-mentioned locality of the interaction, the
kinetic integral $I_2$ can be expanded in a series over the $\gamma_2$
parameter. To the main approximation, it reads \cite{Kolmakov:95a} \be
\dot{n}_k \,=\, \frac{1}{k^2}\, \frac{\prd^2}{\prd k^2 }\, \left( k^{11} \,
n_k^3 \,  \frac{\prd^2}{\prd k^2}\, \frac{1}{n_k} \right).\label{dif} \ee
Equation (\ref{dif}) governs the long-time evolution of the system. It is
similar to the Fokker-Plank kinetic equation for a rarefied gas
\cite{Landau:statphys}, which describes the diffusion of gas density
perturbations. In our case, the perturbations of the distributions of coupled
first and second sound waves diffuse in $K$-space. However, the r.h.s. of Eq.\
(\ref{dif}) is the \textit{fourth order} differential operator and hence it
corresponds to hyperdiffusion. It is worth noting that dynamical equations
similar to (\ref{dif}) arise in the description of nonlinear light pulse
propagation in optical waveguides \cite{Dyachenko:92} and of long  waves on the
ocean surface \cite{Hasselmann:62,Hasselmann:63}.

\subsubsection{Formation of the cascades}\label{sec:nonstationary}
Equation (\ref{dif}) provides a useful tool for the analysis of nonstationary
processes in superfluid systems. In this Section, we focus on the processes
through which the steady-state distributions (\ref{dist}) are formed. We
suppose that the driving force is applied at $t=0$ in a step-like manner.
Assuming self-similarity of the time evolution of the system, one obtains the
instantaneous distribution in the high-frequency spectral domain at times $t>0$
in the following form \cite{Ganshin:10} \be n(\omega,t) = \tau^{q} g(\omega
\tau^{p}).\label{nshort} \ee In Eq.\ (\ref{nshort}) we have used the frequency
representation with $\omega = u_2 k$ to simplify the comparison with
experiment, $g$ is a self-similar function, $q$ and $p$ are positive scaling
exponents, and we denote $\tau = t_0-t$ with the finite formation time $t_0
\sim \tau_2$ where $\tau_2$ is the second sound nonlinear interaction time at
$\omega = \omega_{\rm drive}$. In accordance with the discussion in Sec.\
\ref{sec:diffapp}, we focus on the second sound distribution $n$. The first
sound distribution can be found from an adiabatic relation in each moment of
time. The values of the $q$ and $p$ exponents and the function $g$ can be found
by direct numerical solution of Eq.\ (\ref{dif}). These results will be
discussed in more detail elsewhere.

It follows from Eq.\ (\ref{nshort})  that the formation of the direct cascade
in the high-frequency spectral domain, $\omega>\omega_{\rm drive}$, is of the
``explosion type'' \cite{Zakharov:92}, with a finite formation time $\sim
\tau_2$ for the finite capacity spectrum \cite{Connaughton:10}. The transient
process can be understood as the propagation of a formation front in frequency
space from low to high frequencies, whose position at time $t$ is $\omega_f(t)
\propto (t_0 - t) ^{-p}$.

The formation of the inverse cascade in the low-frequency domain
$\omega<\omega_{\rm drive}$ can be described by the same dependence
(\ref{nshort}), but with different exponents $q$ and $p$ and with $\tau =
\tau_2+t$. In experiments with the superfluid liquid in a cavity, the
accessible range of resonant frequencies is bounded from below by the
fundamental frequency of the cavity, $\omega_0 \sim u_2/L$. If the driving
force is applied at a high resonant frequency, $\omega_{\rm drive} \gg
\omega_0$, the build-up process of the inverse cascade requires a time $t_f
\sim \tau_2 (\omega_{\rm drive} /\omega_0)^{1/p}$ that is much longer than is
needed for formation of the direct cascade. This conclusion is in qualitative
agreement with the experimental observations \cite{Ganshin:08,Efimov:09}.

\subsection{Numerical simulations}\label{sec:numerics}
\subsubsection{Second sound turbulence}
To capture the details of the wave cascades' formation processes, we have
undertaken a numerical study of the dynamics of nonlinear  waves in superfluid
$^4$He within the high-$Q$ resonator. In this investigation, we use parameters
of the model which correspond to those in experiments
\cite{Kolmakov:06,Ganshin:08,Ganshin:10} with superfluid $^4$He. The latter
were conducted at temperatures relatively close to the temperature of the
superfluid transition, $(T_{\lambda} - T)/T_{\lambda} <0.05$. As pointed out in
Sec.\ \ref{sec:represent}, the nonlinearity coefficient $\alpha_2$ grows as $T
\rightarrow T_{\lambda}$ and hence the interaction between the second sound
waves becomes important. On the other hand, the interaction between second
sound and first sound is relatively small in this regime and can to a first
approximation be neglected. Accordingly, the simulations only take account of
terms containing the second sound amplitudes $b_{\bm{k}}$.

The frequency of second sound depends on its wave vector $k$ as \cite{Tyson:68}
\begin{equation}
\omega=u_{20} k \left[1+ \lambda_0 \xi^2 k^2 + \ldots \right],
\label{disp}
\end{equation}
where $\xi= \xi_0 (1- T/T_{\lambda})^{-2/3}$, $\xi_0 \sim 2-3$ A, and
$\lambda_0 \sim 1$. The dispersion of second sound is significant within a
close vicinity of the superfluid transition (i.e.\ for $T_{\lambda}-T<1$
$\mu$K) but is very weak in the temperature range $T<2.1$\,K. However, the
presence of non-zero dispersion is of key importance for the wave dynamics in a
superfluid liquid. It is well established that if the dispersion is exactly
equal to zero, $\partial u_{2}/\partial k = 0$, the main mechanism of energy
transport through the frequency scales is associated with the creation of a
shock wave \cite{Kadomtsev:70}. When the dispersion is positive but small,
$\partial u_{20}/\partial k > 0$ (e.g.\ in the case of second sound in He II),
the resonant three-wave interaction is only relevant within a narrow cone of
${\bf k}$-vectors around the direction of propagation of the wave with ${\bf k}
= {\bf k}_{\rm drive}$ \cite{Zakharov:92,Lvov:00}. The relative phases of waves
with ${\bf k}$-vectors almost collinear to ${\bf k}_{\rm drive}$ (i.e.\ within
the cone) are random, and a kinetic equation for waves can be used to describe
the cascade-like propagation of energy in the wave system
\cite{Zakharov:92,Zakharov:70}. However, the interaction of waves with
non-collinear wave vectors is controlled by higher nonlinear terms and is
relatively small. That is, the wave distribution in a superfluid is nearly
one-dimensional, provided that the dispersion given by Eq.\ (\ref{disp}) is
weak. The formation of this wave regime manifests itself in, for example,
fluctuations of the wave field at high frequencies and in the establishment of
a near-Gaussian probability distribution function for the second sound wave
amplitudes, as was observed in Ref.\ \cite{Efimov:08b}. This regime is a close
analogue of classical wave turbulence \cite{Zakharov:92}. We refer to it as the
\textit{second sound turbulence}.

\begin{figure}[t]
\vspace{3.mm}
\centerline{\includegraphics[width=70.mm,height=50.mm]{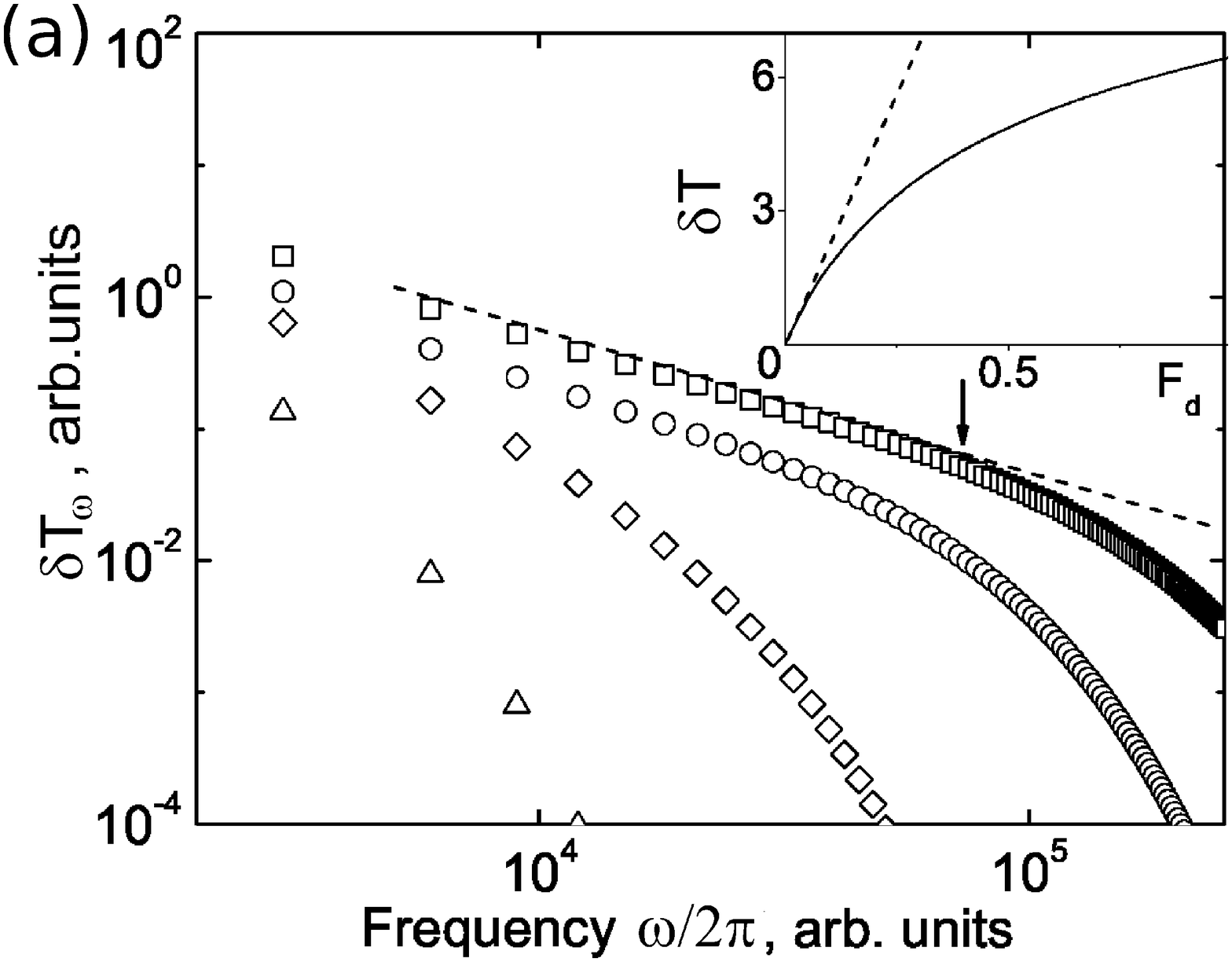}\hspace{0.5cm}
\includegraphics[width=70.mm]{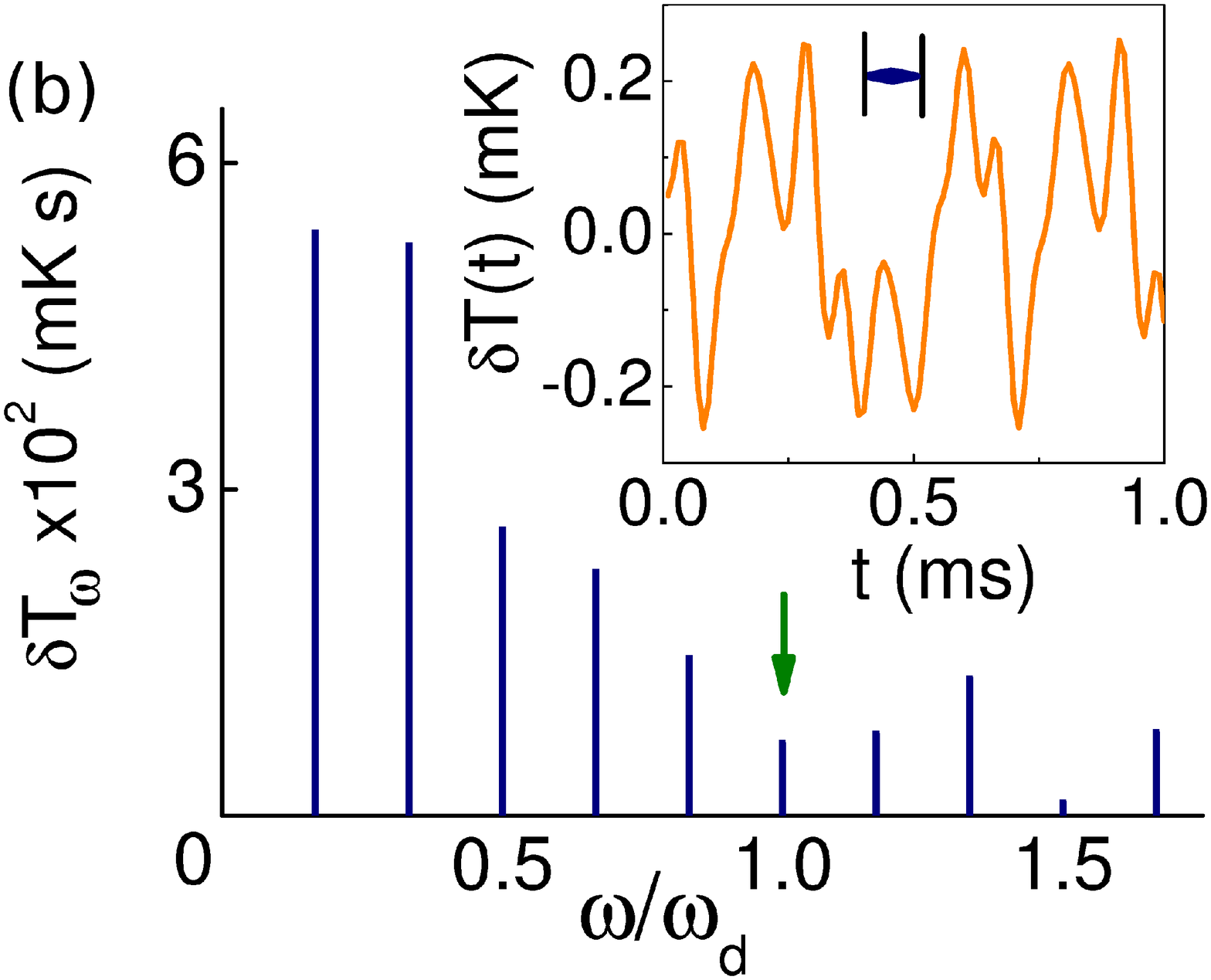}}
\caption{(a) Steady-state second sound power spectral amplitudes
$\delta T_{\omega}$ calculated numerically from Eq.\ (\ref{theoreq1}) for four
different driving force amplitudes: $F_d$ = 0.01 (triangles); 0.05 (diamonds);
0.1 (circles); and 0.3 (squares). The dashed line corresponds to $\delta
T_{\omega} \propto \omega^{-1}$. The arrow marks the boundary frequency
$\omega_b/2 \pi$ of the inertial range for $F_d = 0.3$. The inset shows the
calculated dependence on $F_d$ of the standing wave amplitude $\delta T$: solid
line -- nonlinear waves, $\alpha_2 <0$ ($T>T_{\alpha}$); dashed line -- linear
waves, $\alpha_2 =0$ ($T=T_{\alpha}$). (From Ref. \cite{Kolmakov:06}.)
(b) Formation of the inverse cascade of second sound waves spectrum
computed for $t=1000$. Main plot: the wave spectrum in presence of the inverse
cascade computed for $N=10$ modes in the resonator. The vertical arrow
indicates the frequency $\omega_{\rm drive}$ of the driving force. Inset:
oscillations of temperature at a wall of the resonator. The horizontal arrow
shows the driving force period.} \label{fig:steadystate}
\end{figure}

The nonlinear dynamics of second sound involves the interactions of sound waves
on different scales. Hence the full equations (\ref{abeqs}) should be solved
numerically. In these simulations, we consider one-dimensional longitudinal
second sound waves in a long cylindrical resonator \cite{Kolmakov:06}. We
neglect the possible generation of transverse, radial modes. Note, however,
that the excitation of transverse modes in a cylindrical resonator can be
important for non-planar, three dimensional waves \cite{Zinovieva:57}. The
longitudinal resonant wave vector is $k_n = \pi n /L$ where $L$ is the length
of the resonator and $n$ is the number of the resonant wave mode. The equations
of motion (\ref{abeqs}) governing non-local energy balance in the second sound
system read
\begin{equation}
 i {\partial {b}_n \over \partial t} = \sum_{n_1, n_2} V_{n,n_1,n_2}
 (b_{n_1} b_{n_2} \delta_{n-n_1 -n_2} + 2 b_{n_1} b_{n_2} ^* \delta _{n_1 - n_2 -n} )
 - i \gamma_n b_n + F_{d}. \label{theoreq1}
\end{equation}
Here, $V _{n,n_1,n_2} \propto \alpha_2 (n n_1 n_2)^{1/2}$ describes the
three-wave interaction, $\gamma_n = \nu n^2$ models the viscous damping of
second sound, $F_{d} \propto W $ is the amplitude of the force driving the
$n$-th resonant mode \cite{Brazhnikov:04a,Kolmakov:06}. In the present model,
the wave damping was taken into account at all frequencies, a feature that is
of key importance \cite{Kolmakov:06} for a correct description of the formation
of a cascade of nonlinear sound waves with increasing amplitude of the driving
force. The wave spectrum is calculated as $\delta T_{\omega} = B_n \, (b_n +
b_n^*)$ where $B_n = (\omega_n /L)^{1/2} (\partial S/ \partial T)^{1/2}$. We
use the dispersion relation for linear waves, $\omega_n = u_2 k_n$, to compute
the resonant frequencies of the cavity.

Fig.\ \ref{fig:steadystate}(a) shows the evolution of the steady-state spectrum
with increasing driving force amplitude $F_{d}$, calculated for $\alpha_2 < 0$
($T>T_{\alpha}$). We used a periodic driving force of frequency equal to a
resonant frequency of the resonator corresponding to the conditions of the
measurements, taken as $\omega_{\rm drive}/2 \pi=3000$ in dimensionless units.
The effective viscosity coefficient $\nu$ was fitted to the measured value of
the quality factor $Q \sim 3 \times 10^3$ of the resonator, to facilitate
comparison of the model results with those from the experiments
\cite{Kolmakov:06}. Points on the plot correspond to the amplitudes of the
peaks in the spectrum. It is seen that, at small driving amplitude  $F_d \sim
0.01$ (triangles), viscous damping prevails at all frequencies  and  a
turbulent  cascade is not formed: the amplitude of the second harmonic is an
order less that the amplitude of the main harmonic. In this regime the wave
shape is close to linear. At intermediate driving amplitude $F_d \sim 0.05$
(diamonds) nonlinearity starts to play a role at frequencies of the order of
driving frequency, and a few harmonics are generated. At high driving
amplitudes $F_d \geq 0.1$ (circles and squares) a well-developed cascade of
second sound waves is formed up to frequencies 30 times higher than the driving
frequency.

\subsubsection{The inverse cascade}
To capture the main characteristic features of the development of the inverse
cascade while, at the same time, keeping the system simple enough to be
analyzed in detail, we carry out a simulation within the framework of equations
(\ref{theoreq1}) taking account of the relatively small numbers of waves $N$:
the summation on the r.h.s.\ of Eq.\ (\ref{theoreq1}) was made for $n_1$, $n_2$
ranging from 1 to $N$. In what follows we present the results obtained for
$N=10$. We assume that a periodic driving force is applied at the 6th resonant
frequency.

It is was observed in simulations that, for sufficiently high driving amplitude
$W$, low frequency harmonics at $\omega < \omega_{\rm drive}$ are formed in the
wave spectrum (see Fig.\ \ref{fig:steadystate} (b)). Consistent with the
discussion above, this is a manifestation of the formation of an inverse
cascade in a superfluid. The numerical results describe very well the data
obtained in the experiments on the inverse cascade of second sound waves in
He-II (see Sec.\ \ref{sec:expinverse} below).

It is clearly evident from Fig.\ \ref{fig:steadystate}(b) that the
inverse cascade is responsible for the formation of high-amplitude,
low-frequency, subharmonics (cf.\ the experimental results shown in Fig.\
\ref{fig:exp1}(b)). As it is shown in Refs.\
\cite{Ganshin:08,Efimov:10}, the inverse cascade develops through formation of
isolated low-frequency waves of higher amplitude than is typical of the waves
around them. These higher-amplitude lone waves can be considered as the
acoustic analogue of the giant ``rogue'' waves that occasionally appear on the
ocean and endanger shipping. Their origin lies in a decay instability of the
periodic wave, i.e.\ a similar mechanism to that proposed
\cite{Onorato:01,Dyachenko:05} (modulation instability) to account for the
creation of oceanic rogue waves \cite{Dean:90}.

\begin{figure}[t]
\begin{center}
\includegraphics[width=70mm]{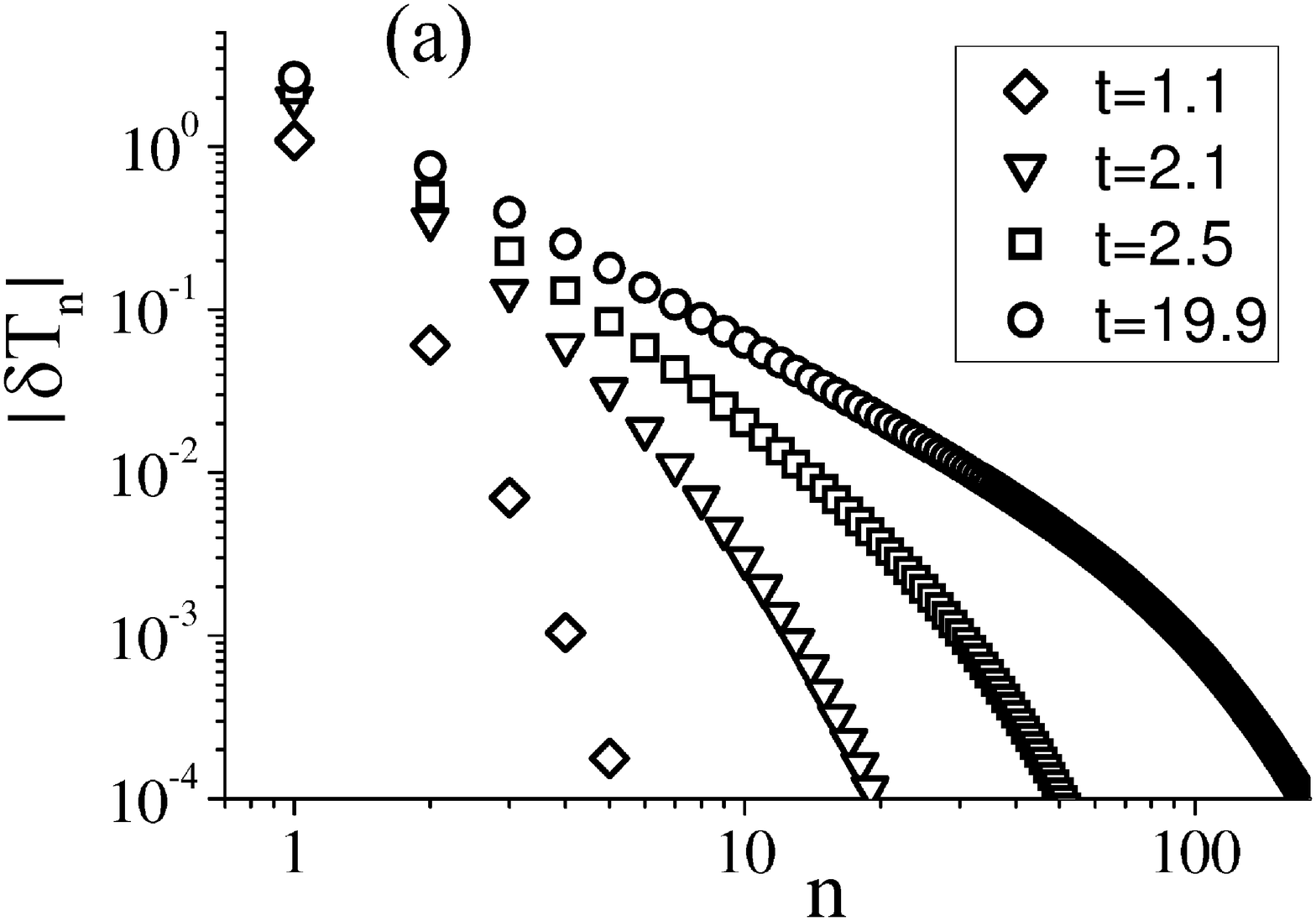}
\includegraphics[width=70mm]{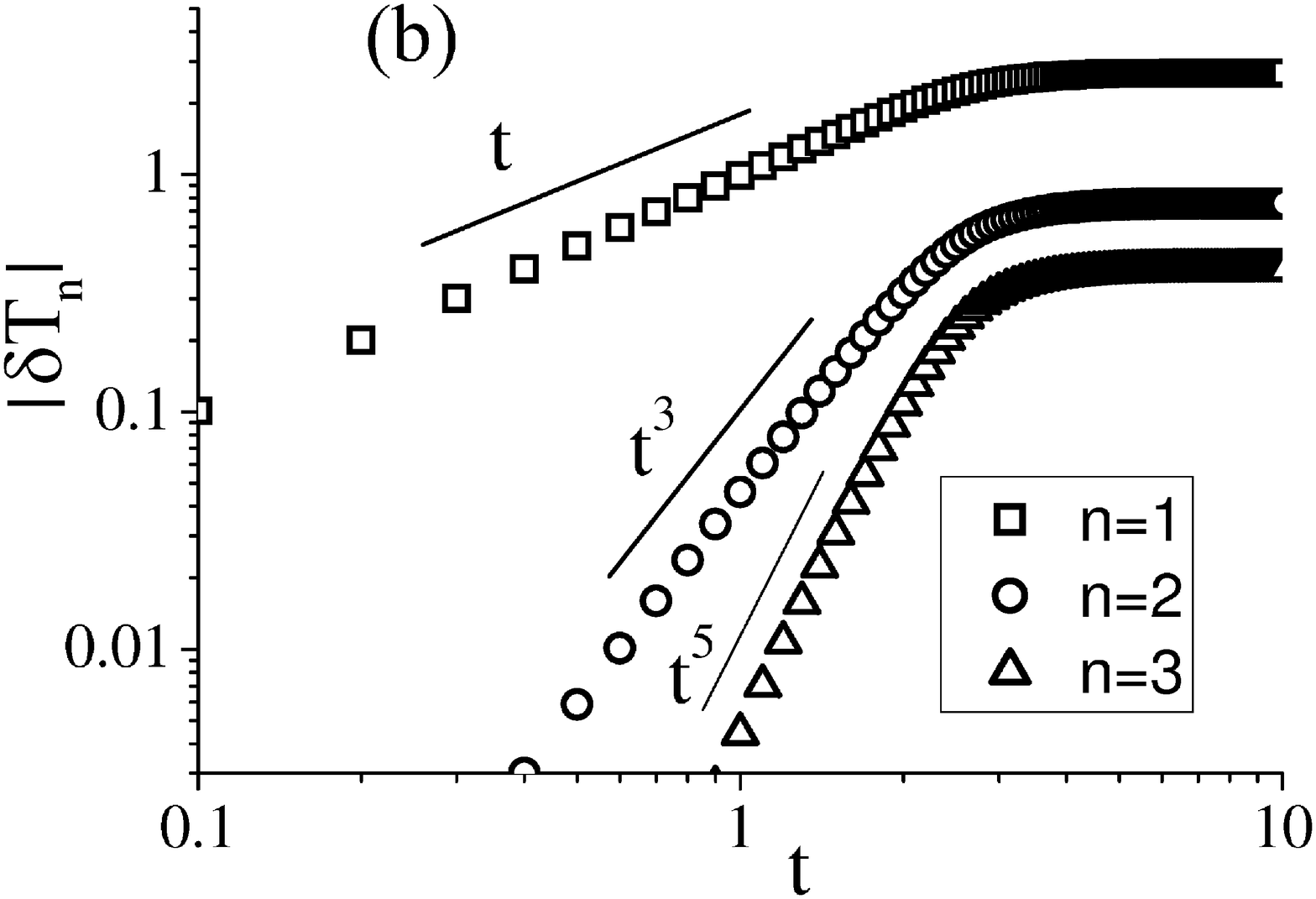}
\vspace{-0.2cm}
\caption{(a) Evolution with time $t$ of the spectral amplitude during the
build-up of second sound turbulence. (b) Points: dependencies on $t$ of the
spectral amplitude at driving frequency, $\delta T_1$, and of higher harmonics,
$\delta T_n$ with $n>1$. Lines: dependencies $\delta T_n \propto t^{2n-1}$. The
computations were for $F_d=1$. From Ref. \cite{Ganshin:10}} \label{numer}
\end{center}
\end{figure}

\subsubsection{Turbulence formation}
To capture the dependence on time of the formation of the direct cascade, we
compute the evolution of the second sound wave spectrum with time for the
driving force $F=1$ (in numerical units). The steady-state spectrum established
at large times, i.e., after all transient processes are finished at $t>t_{form}
\sim 5$, is shown in Fig.\ \ref{numer}(a) by circles. The numerical unit of
time we use in simulations is equal, in order of magnitude, to the
characteristic time $\tau_2$ taken at the driving frequency. Figure
\ref{numer}(a) also demonstrates the evolution of the spectrum during the
build-up process. Figure \ref{numer}(b) shows the dependence on time of the
amplitude of the wave, $\delta T_n$ at the driving frequency ($n=1$) and of two
higher harmonics at $n=2$ and 3, computed for the evolution presented in Fig.\
\ref{numer}(a).  (We recall that the second sound oscillations observed in
experiments are fluctuations in the temperature of the superfluid). It is
clearly evident that the dependence of the wave amplitudes on time during the
formation process $t<2$ is well described by a power-law function $\delta T_n
\propto t^m$, where $m=2n-1$. Note that this power-law dependence has also been
found in analytical computations through a small-time expansion in Eqs.\
\ref{theoreq1} \cite{Ganshin:10}.

The nonstationary process through which second sound turbulence is formed has
recently been investigated in experiments Ref. \cite{Efimov:10}. It was found
that the turbulence build-up process is well described by the self-similar
dependence (\ref{nshort}) and is in agreement with Fig.\ \ref{numer}. From a
comparison of experimental data with Eq.\ (\ref{nshort}) the exponent was
estimated to be $p = 5$.

\subsection{Experimental observation of the direct and the inverse cascades}\label{sec:exp}
\subsubsection{Experimental arrangements}\label{sec:expdetails}
\begin{figure}[t]
\begin{center}
\includegraphics[width=80.mm]{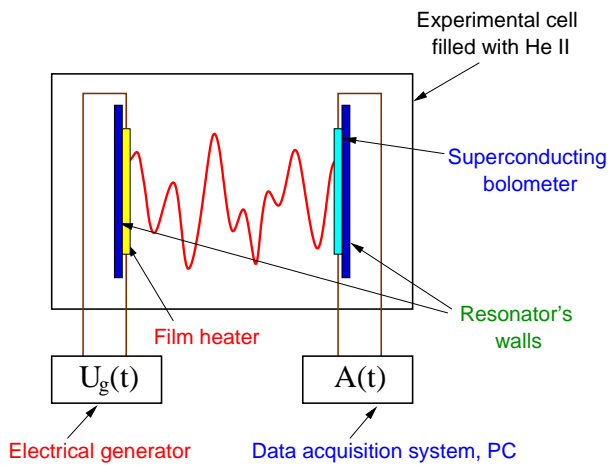}
\caption{Schematic diagram illustrating the experimental
arrangements. The cylindrical walls of the resonator are not
shown. From Ref.\ \cite{Efimov:09}}
\label{fig:expscheme}
\end{center}
\end{figure}

Detailed experimental investigations of second sound turbulence in superfluid
helium in a high-quality resonator have recently been undertaken
\cite{Kolmakov:06,Efimov:06,Efimov:07,Ganshin:08,Ganshin:08b,Efimov:08,Ganshin:10}.
In these studies, a thin-film heater was utilized to generate the second sound,
and a thin-film bolometer (a fast-acting superconducting thermometer) was used
as a detector: see Fig.\ \ref{fig:expscheme}. The use of a high-$Q$ resonator
enables one to create nonlinear standing second sound waves of high amplitude
with only small heat production at the source. The resonator was formed by a
cylindrical quartz tube of length $L=7$ cm and inner diameter $D=1.5$ cm. The
film heater and bolometer were deposited on the surfaces of flat glass plates
capping the ends of the tube. The heater was driven by an external sinusoidal
voltage generator in the frequency range between 0.1 and 100 kHz: The frequency
of the second sound (at twice the frequency of the voltage generator) was set
equal to one of the longitudinal resonant frequencies.

\subsubsection{The direct cascade of second sound turbulence}
For a small ac heat flux density $W<1$ mW/cm$^2$, corresponding to a standing
wave amplitude less than 0.5 mK, a nearly linear regime of wave generation was
observed: the amplitude $\delta T$ of the standing wave was proportional to the
heat flux density $W$. Increase of the excitation above a few mW/cm$^2$ led to
large deviations from the linear dependence $\delta T \propto W$, however, and
to visible deformation of the initially sinusoidal standing wave, accounted for
by the formation of multiple harmonics in its spectrum. Fig.\ \ref{fig:exp1}(a) 
shows some typical experimental results at 2.079~K. The formation
of multiple harmonics in the spectrum corresponds to the establishment of
second sound turbulence in superfluid helium in the resonator.

It is clearly evident from Fig.\ \ref{fig:exp1}(a) that the main
spectral peak lies at the driving frequency $\omega_{\rm drive}$, and that
high-frequency peaks appear at its harmonics, $\omega_n=\omega_{\rm drive}
\times n$ with $n=2,3,\ldots$ It can be seen that a cascade of waves is formed
over the frequency range up to 80 kHz, i.e.\ up to a frequency 25 times higher
than the driving  frequency. The dependence of peak height on frequency may be
described by a power law  $\delta T_{\omega} = {\rm const} \times \omega_n
^{-s}$ at frequencies lower than  some cutoff  frequency $\omega_b/2 \pi \sim
5.5 \times 10^4$ Hz. For sufficiently high heat flux densities $W>10$ mW/cm$^2$
the scaling index tends to $s\approx 1.5 \pm 0.3$.

\subsubsection{Build-up of the inverse cascade}\label{sec:expinverse}
\begin{figure}[h]
\centerline{\includegraphics[width=75.mm]{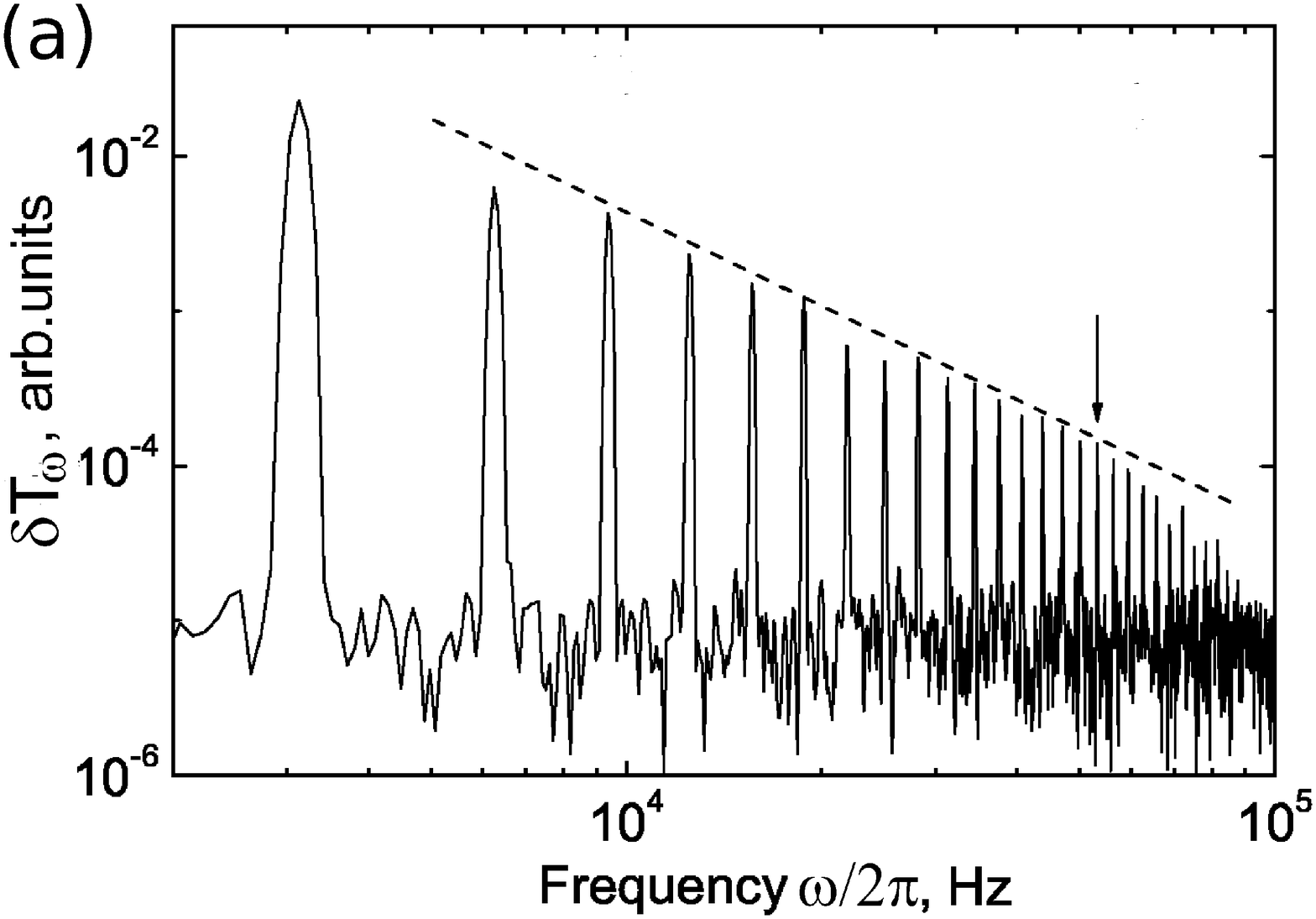}
\includegraphics[width=70.mm]{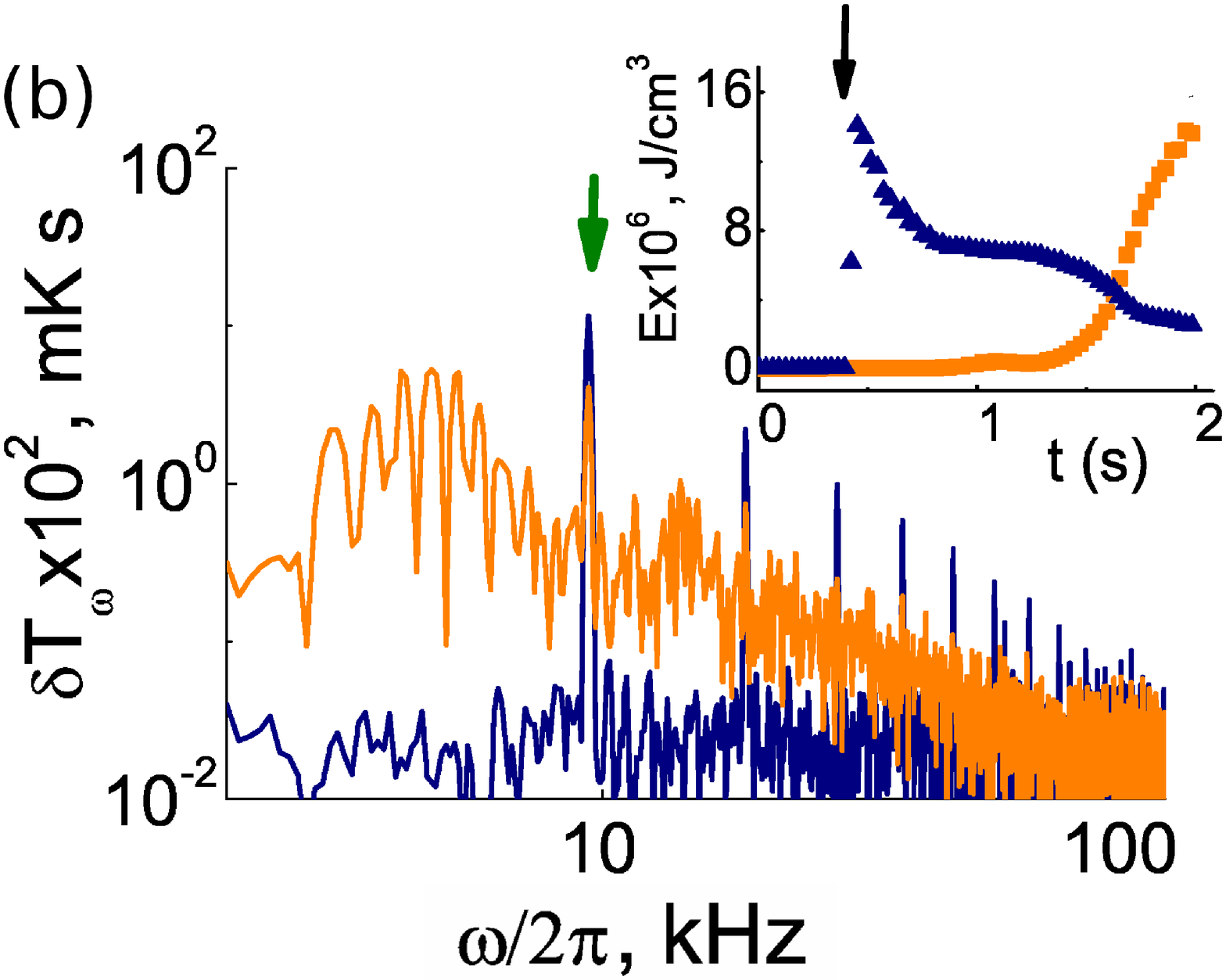}}
\caption{(a) Power spectral amplitudes $\delta T_{\omega}$ of standing waves
recorded at $T=2.079$ K when driving at the 31st resonant
frequency, $\omega_{\rm drive}/2 \pi=3130$ Hz. The ac heat flux density from the heater
was 22 mW/cm$^2$. The dashed line corresponds to $\delta T_{\omega} \propto \omega^{-1.5}$. The
arrows indicate  positions of the viscous cutoff frequency.
(b)  Formation of the inverse cascade of second sound turbulence.
The lower blue spectrum shows the direct cascade only; the upper orange
spectrum shows both the direct and inverse cascades formed at later stage
of the turbulence formation. The green arrow indicates the fundamental peak at the
driving frequency. Inset: evolution of the wave energy in the
low-frequency and high frequency domains is shown by the orange
squares and blue triangles respectively; black arrow marks the
moment of time at which the external drive was switched on.
From Refs. \cite{Kolmakov:06,Ganshin:08}} \label{fig:exp1}
\end{figure}

Formation of the inverse cascade of second sound turbulence was observed when a
high enough periodic heat flux density was applied at a small detuning from the
resonance. It was found that in this case turbulence formed in two stages.
First, the direct cascade developed after the heat flux was switched on, as
shown by the blue spectrum in Fig.\ \ref{fig:exp1}(b). The green arrow
on the main plot marks the driving frequency $\omega_{\rm drive}$. The moment
$t = 0.397$ s, at which the the driving voltage was switched on, is labeled by
a black arrow in the inset. At the second stage, subharmonics of the driving
frequency at $\omega < \omega_{\rm drive}$ were generated, as demonstrated by
the orange spectrum on Fig.\ \ref{fig:exp1}(b). Orange triangles in
the inset show the dependence on time of the wave energy accumulated in the
low-frequency region $\omega < \omega_{\rm drive}$. The reduction of wave
amplitude seen in the high frequency spectral domain (blue triangles) is
indicative of the onset of energy backflow towards lower frequencies, i.e.\ a
sharing of the input energy between the direct and inverse cascades. The
decrease in energy at high frequencies in $0.397 s < t < 1.3$\,s is
attributable to relaxation processes in the direct cascade. Redistribution of
wave energy due to sharing of the energy flux between the direct and inverse
cascades starts at $t=1.3$\,s. The results of our numerical computations
discussed above in Sec.\ \ref{sec:numerics} are in agreement with these
experimental observations.

\section{Bose-Einstein condensation and superfluidity of
polaritons in a microcavity}\label{sec:polaritons}
\subsection{Microcavity polaritons}\label{sec:bec}

\begin{figure}
\begin{center}
\includegraphics[width=60.mm]{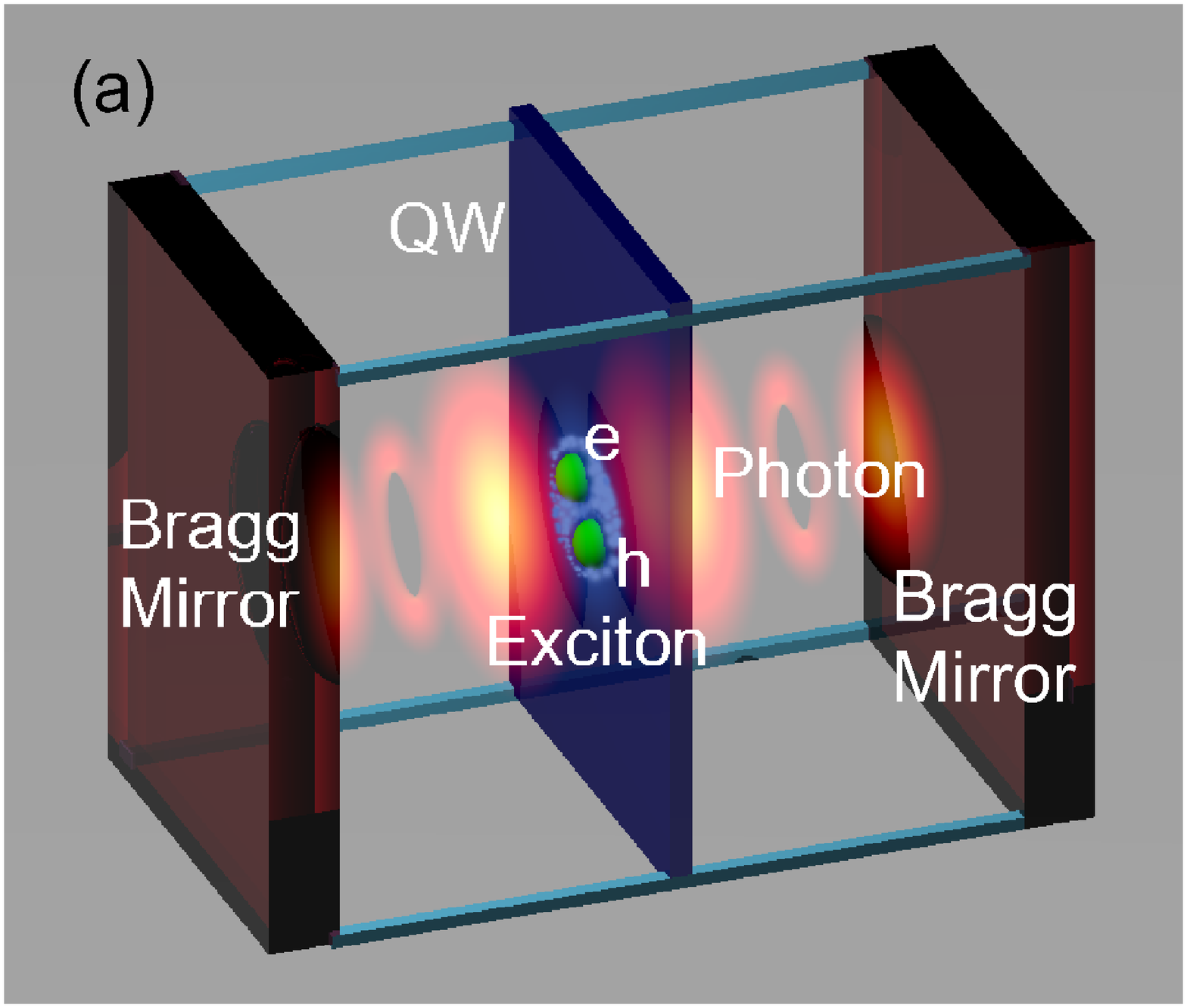}\hspace{0.5cm}
\includegraphics[width=70.mm,height=45.mm]{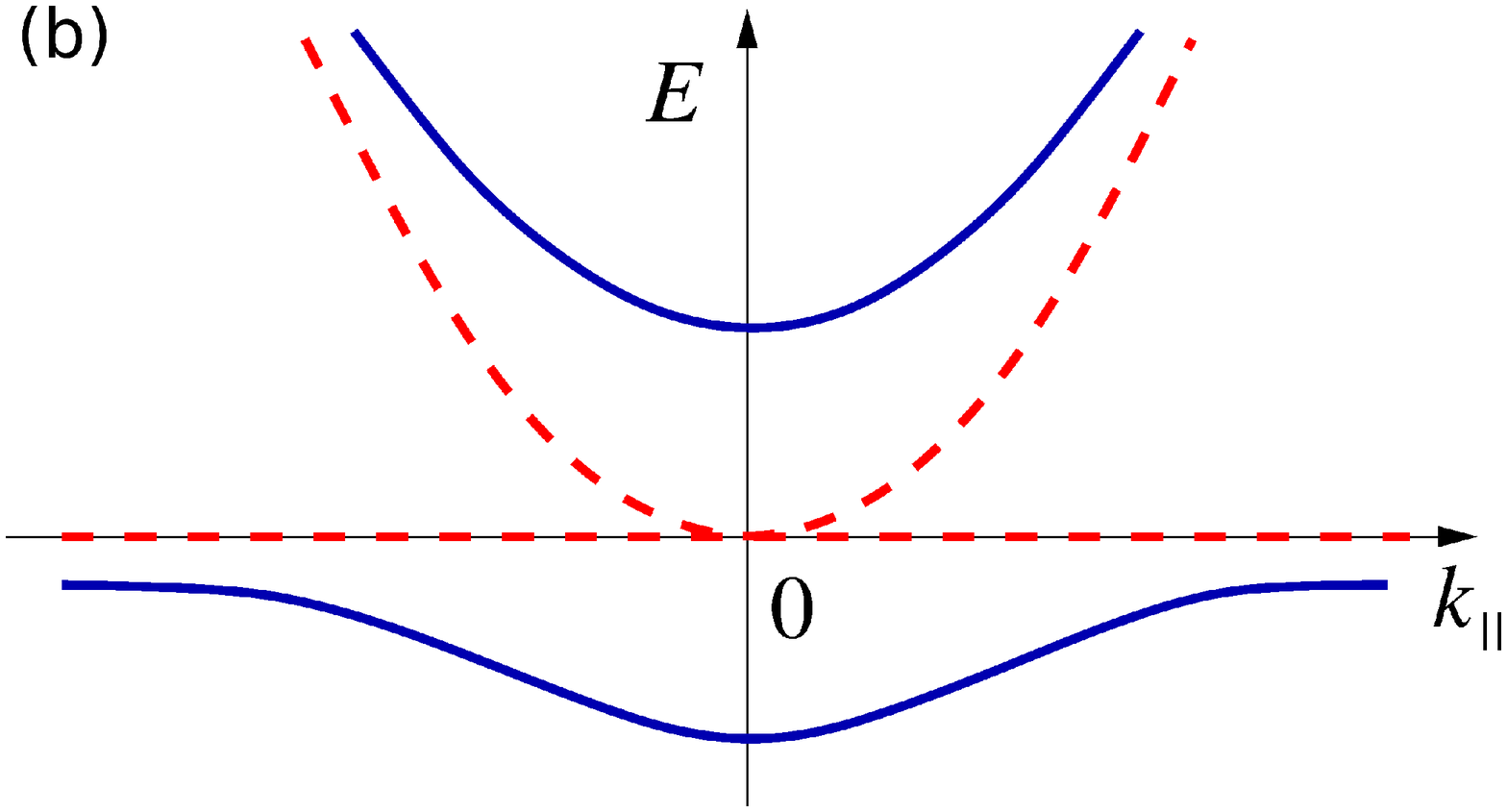}
\end{center}
\caption{(a) Exciton in an optical microcavity: a quantum well is placed
between two mirrors composed of multilayer Bragg
reflectors. (b) Parabolic red dash curve: dispersion of photons in a planar
microcavity, $k_{||}$ is a component of the photon wave vector parallel to the
mirrors. Horizontal red dash line: Energy of excitons in a quantum well
resonant with the cavity photon energy. Compared to the dispersion
of the cavity photons, the exciton energy is essentially constant
near $k_{||} = 0$. Solid blue lines: upper and lower polariton modes
formed by anticrossing of the photon and exciton states.} \label{cav}
\end{figure}

In this Section, we consider another important example of a quantum system, in
which the Bose-Einstein condensation occurs at low enough  temperatures -- the
exciton polaritons in a quantum well (QW) embedded into an optical microcavity.
The microcavity just consists of two mirrors, which are built opposite to each
other in a semiconductor heterostructure ~\cite{Snoke,Snoke_text,Lit}, as shown
in Fig.~\ref{cav}(a). The two mirrors are typically fabricated as sets of
dielectric layers forming a Bragg reflector~\cite{Snoke}. The cavity is open in
the transverse direction, so that light is free to move out of the cavity if it
propagates parallel to the mirrors, but it is confined to specific resonant
modes in the longitudinal direction perpendicular to the mirrors. The quantum
well is placed in between these two mirrors, as seen in Fig.~\ref{cav}(a). The
\textit{excitons}, which are the bound states of an electron in the conduction
band coupled to a hole in the valence band, are created by external laser
pumping \cite{Snoke_text}. Let us note that by designing Bragg reflectors of
proper thickness, it is possible to tune the fundamental frequency of the
cavity to be equal to the energy of the excitons; under these conditions light
and excitons interact resonantly because of the non-zero electric dipole moment
of the excitons, thus forming two new branches of excitations. These new
branches, known as the upper and lower \textit{polaritons}, are superpositions
of the excitons and  photons in the microcavity (see Fig. ~\ref{cav}(b))
~\cite{Hopfield,Kavokin_book,Snoke_text}.

The effective mass of the lower polariton branch is given by the curvature of
the band at zero wave vector in the longitudinal direction of the QW, i.e.\ at
$k_{||} = 0$, where $k_{||}$ is a component of the photon wave vector $k$
parallel to the Bragg mirrors. Because of the small curvature of the band, the
polaritons are extremely light relative to atoms and excitons (cf.\ discussion
and references below). Under typical experimental conditions, the effective
mass of a polariton can be as low as $10^{-4} \times m_0$, where $m_0$ is  the
free electron mass~\cite{Snoke_text}. Consequently, as we demonstrate below,
the onset of BEC and superfluidity occurs in the system of polaritons at much
higher temperatures than in systems of atoms or
excitons~\cite{Lit,Snoke_2006,Snoke_2008}.

It is worth noting that the quantum well in an optical microcavity behaves like
a laser because the light emitted from the polariton  condensate is nearly
coherent~\cite{Lit}. However, this special kind of  laser does not require an
inverse population of excitons and thus, theoretically, it has no
threshold~\cite{Kavokin_laser}. Several  recent
experiments~\cite{Kasprzak:06,Amo:09} have demonstrated spontaneous coherence
in exciton-polariton systems in various two-dimensional (2D) semiconductor
microcavity structures. In addition, the resonant laser pumping can promote, in
its turn, condensate formation as a result of nonlinear light
scattering~\cite{Stevenson}.

\subsection{Bose-Einstein condensation and superfluidity of two-dimensional
polaritons in an in-plane harmonic potential}

The properties of polaritons have been studied in several theoretical works.
The theory of polariton dynamics due to the polariton-polariton interaction has
been developed in Refs.~\cite{Ciuti-exex,Tassone,Porras}. The crossover between
lasing and polariton coherence has been studied in
Refs.~\cite{Eastham,keeling}. Polaritons superfluidity and spontaneous linear
polarization of the light emission were predicted in Refs.
\cite{Carusotto:04,kav-lin}. In these earlier studies, the coherent polaritonic
phases were analyzed in unrestricted 2D system.

At finite temperatures, $T>0$, there is no BEC in an infinite two-dimensional
(2D) system \cite{Landau:statphys}. Nonetheless, a BEC phase transition can be
obtained in a finite-sized system in the presence of a confining potential
\cite{Bagnato,Nozieres}. Recently, polariton systems in a harmonic potential
trap have  been studied experimentally in a GaAs/AlAs quantum well embedded in
a GaAs/AlGaAs microcavity \cite{Balili_apl}. For this kind of trap, the exciton
energy was tuned by applying stress to the host material, thus changing the
inter-band gap energy. A BEC of polaritons in a quantum well has been observed
experimentally in such systems \cite{science_b}.

Below, we briefly review the theory of Bose-Einstein condensation in the
polariton system \cite{Berman:08a}. In the computations, we take advantage of
the Hamiltonian approach. As it was pointed above, in a quantum system the
Hamiltonian normal variables should be changed to operators that act in the
occupation number space \cite{Landau:statphys,Abrikosov}. The operators of
creation and annihilation of polariton in a lower energy band,
$\hat{\psi}^{\dagger}(\mathbf{r})$ and $\hat{\psi}(\mathbf{r})$, are defined as
\begin{equation}
 \psi({\bf r}) = \sum_{\mathbf{P}} \hat{p}_{\mathbf{P}} e^{i{\mathbf{P}\mathbf{r} \over \hbar}},
 \qquad \psi^{\dagger}({\bf r}) = \sum_{\mathbf{P}} \hat{p}_{\mathbf{P}}^{\dagger}
 e^{-i{\mathbf{P}\mathbf{r}\over \hbar}}  \label{pol_op}.
\end{equation}
Here, we use an expansion over running plane waves, $\hat{p}_{\mathbf{P}}$ and
$\hat{p}_{\mathbf{P}}^{\dagger}$ that are the annihilation and creation
operators of polaritons with the momentum $\mathbf{P} = \hbar \bm{k}$. The latter
are the quantum analogues of the normal coordinates $a_{\bm{k}}$ and
$a_{\bm{k}}^*$ at the wave vector $\bm{k}$. The operators
$\hat{\psi}^{\dagger}(\mathbf{r})$ and $\hat{\psi}(\mathbf{r})$ taken at the
same moment of time $t$ satisfy the Bose commutation
relations~\cite{Abrikosov,Fetter}.

As we note above, we consider a ``cloud'' of excitons trapped in a
semiconductor potential well, described by the function $V(r)$ with $r$ being
the distance from the centre of the potential. The external potential $V(r)$
can arise, for example, from a shift of the exciton energy by application of
inhomogeneous stress to the host material~\cite{Balili_apl}. In our model, we
also suppose that the photon states in the cavity are unaffected by the stress.
In the computations,  we use the simplest, harmonic approximation for the
potential energy in the form $V(r) = (1/2) \gamma r^{2}$; the parameter
$\gamma$ characterizes the strength of the interaction. Below we also refer to
$\gamma$ as a ``spring constant'', in analogy with the classical relation
between the energy of a linear spring and the displacement $r$.

In the case where that both the polariton momentum $P$ and the spring
constant $\gamma$ are small the effective Hamiltonian for the lower-band
polaritons in the parabolic trap $V(r)$ acquires the following
form~\cite{Berman:08a}:
\begin{equation}
\label{Ham_eff_real} \hat H_{\rm eff}  = \int d\mathbf{r} \
\hat{\psi}^{\dagger}(\mathbf{r})\left( - \frac{\hbar^2 \Delta}{2M_{\rm eff}}
+ V_{\rm eff}(r) \right)\hat{\psi}(\mathbf{r}) +
\frac{U_{\rm eff}^{(0)}}{2} \int d\mathbf{r} \
\hat{\psi}^{\dagger}(\mathbf{r})
\hat{\psi}^{\dagger}(\mathbf{r})\hat{\psi}(\mathbf{r})\hat{\psi}(\mathbf{r}),
\end{equation}
where the effective mass of a polariton is defined as
\begin{eqnarray}
\label{Meff} M_{\rm eff}^{-1} = \frac{1}{2}   \left(M^{-1} +
(c/n)\lambda_C/(2\pi \hbar) \right) \ .
\end{eqnarray}
Here $c$ is the speed of light, $n = \sqrt{\epsilon}$ is the refraction index
in the microcavity, $\epsilon$ is the dielectric constant, and $\lambda_C$ is
the length of the microcavity (i.e.\ the spacing between the Bragg mirrors), $M
= m_e+m_h$ is the exciton mass, $m_e$ and $m_h$ are the effective masses of
electron and hole.

The effective external potential is equal to $V_{\rm eff}(r) = \frac{1}{2}
\gamma_{\rm eff} r^{2}$  with $\gamma_{\rm eff} =  \gamma/2$. The last term in
Eq.~(\ref{Ham_eff_real}) describes the interaction of polaritons (a quantum
analogue of the four-wave scattering considered in Sec.\ 2).  The
polariton-polariton interaction is purely repulsive, with the hard-core contact
potential $U_{\rm eff}(\mathbf{r} - \mathbf{r}') = U_{\rm eff}^{(0)}
\delta(\mathbf{r} - \mathbf{r}')$, where $U_{\rm eff}^{(0)}=
3e^{2}a_{2D}/(2\epsilon )$ is the  effective interaction constant, $e$ is the
charge of electron, $a_{2D} = \hbar^{2}/(2m_{e-h}e^{2})$ is the two-dimensional
Bohr radius of the exciton, and $m_{e-h} = m_{e}m_{h}/(m_{e}+m_{h})$ is the
exciton reduced mass.

As can be seen from Eq.\ (\ref{Ham_eff_real}), the effective Hamiltonian for
the trapped polaritons maps onto the Hamiltonian of a weakly-interacting 2D
Bose gas in the confining parabolic trap $V_{\rm eff}(r)$. As it shown in Ref.
\cite{Bagnato}, in a parabolic trap the Bose particles can form a Bose-Einstein
condensate, and hence enter a superfluid state at temperatures below a critical
temperature $ T_{c}^{(0)}= \hbar k_{B}^{-1} \sqrt{3 \gamma_{\rm eff} N/ \pi
M_{\rm eff}}$, where $k_{B}$ is Boltzmann's constant. According to Eq.\
(\ref{Ham_eff_real}), this expression gives an estimate for the temperature of
the BEC transition, $T_{c}^{(0)}$ for weakly interacting polaritons.

In this approach, we can  estimate the number of polaritons in the condensate
fraction at $T<T_{c}^{(0)}$ assuming that the polariton system is in thermal
equilibrium. The approximation is valid if the characteristic time for
polariton-polariton scattering is short compared to the polariton lifetime, in
which case the polariton gas rapidly thermalizes and can therefore be
characterized by well-defined thermodynamic temperature. Note that, this
polariton temperature can be substantially higher than the lattice temperature
of the host semiconductor sample. Although the polariton lifetime is typically
short, $\sim 2$ ps, the polariton-polariton scattering time is even shorter at
sufficiently high polariton density. A recent study \cite{Kasprzak:06} has
demonstrated that, in an actual experimental system, this requirement is
satisfied and polaritons are indeed characterized by a well-defined
temperature.

\begin{figure}[t]
\begin{center}
\includegraphics[width=75.mm]{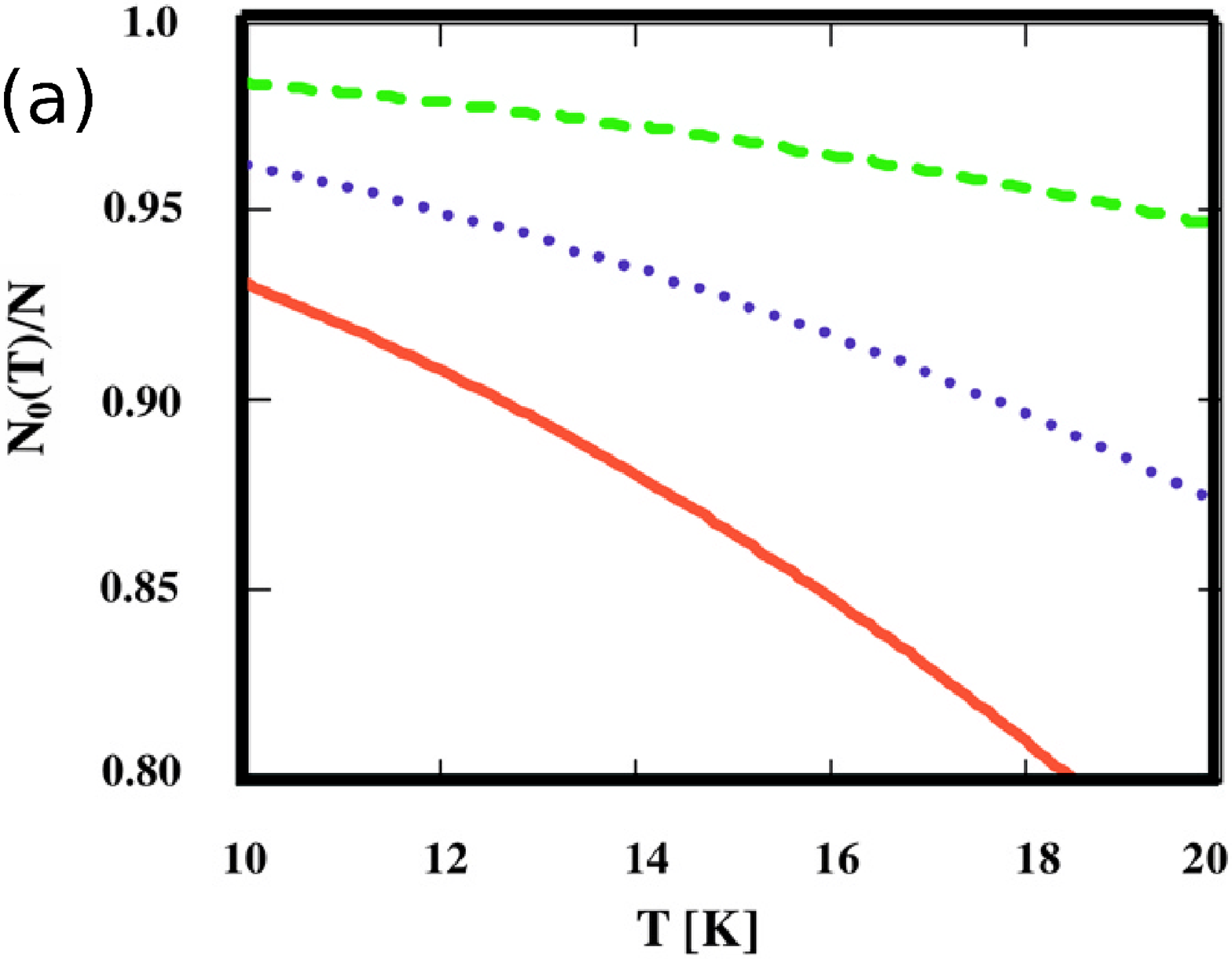}
\includegraphics[width=70.mm]{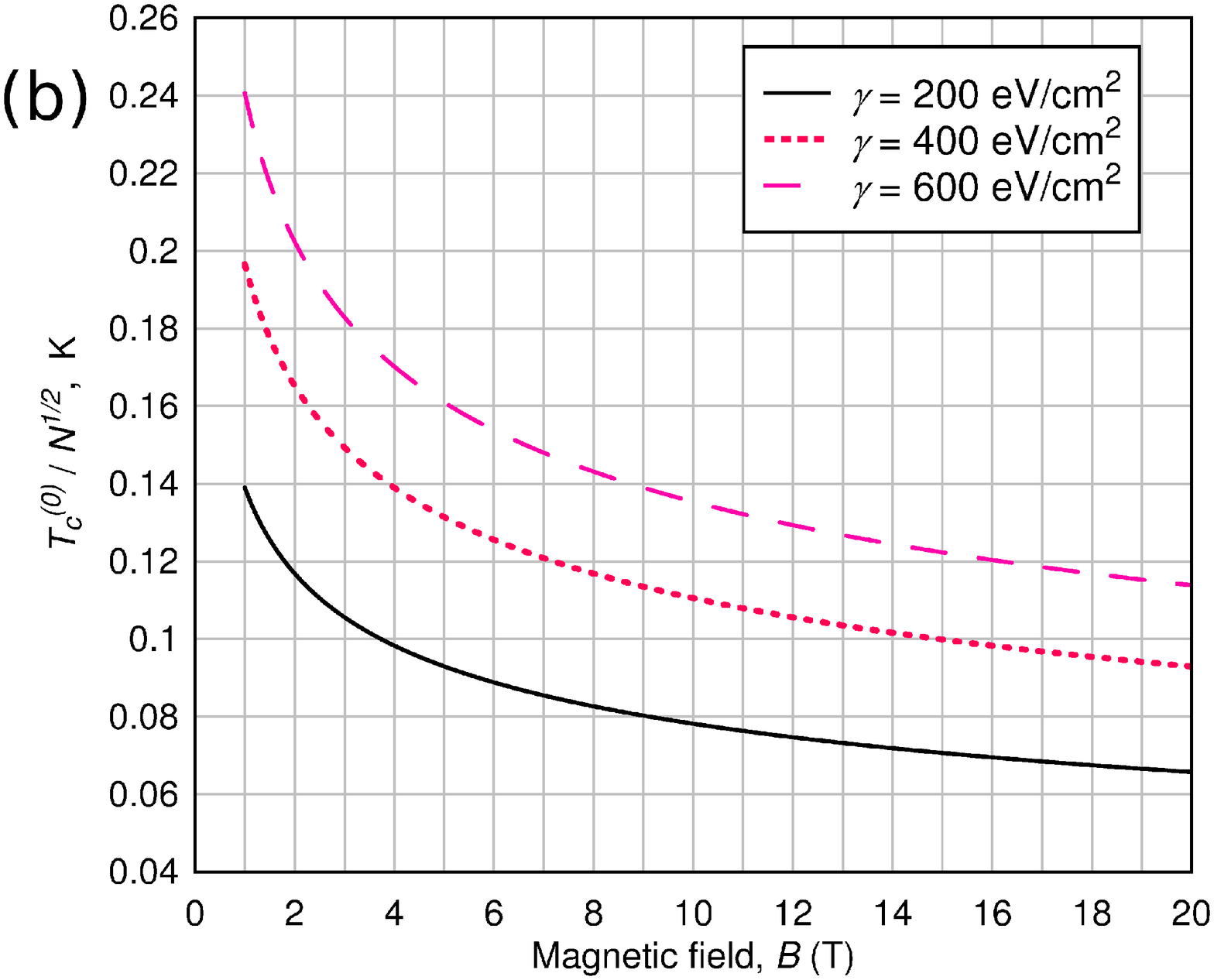}
\end{center}
\caption{(a) Condensate fraction $N_0/N$ of polaritons
in a parabolic trap in an optical microcavity as a function of temperature $T$,
calculated for three values of trapping potential strength,
(solid line) $\gamma = 760$ eV/cm$^2$; (dotted line) 860 eV/cm$^2$ and
(dashed line) 960 eV/cm$^2$. (b) Ratio of BEC critical
temperature to the square root of the total number of
magnetopolaritons $T_{c}^{(0)}/\sqrt{N}$ as a function
of magnetic field $B$  at different trapping potential strength
$\gamma$. We assume that the environment around the graphene layer
is $\mathrm{GaAs}$ with $\epsilon = 12.9$. From Refs.
\cite{Berman:08a,Berman_magnetopolaritons}} \label{N0}
\end{figure}

Below we restrict our consideration to the range of temperatures below
$T_{c}^{(0)}$ but effectively above the lowest energy level in the confining
potential, $T \gg \hbar \omega_0 /k_{B}$, where $\omega_0 = (\gamma_{\rm
eff}/M_{\rm eff})^{1/2}$. For example, by taking $\gamma = 100 \ \mathrm{
eV/cm^{2}}$ as an estimate for typical experimental conditions, this implies $T
\gg 0.3$ K. The latter condition allows us to neglect geometrical quantization
of the polariton energies in the confining potential, and hence use the
quasiclassical approach to the problem \cite{Berman:08a}. Figure~\ref{N0}({\it
b}) shows the fraction $N_{0}(T)/N$ of polaritons in the condensate at a given
temperature $T$  for the experimental parameters provided in
Ref.~\cite{Berman:08a}. The computations were done  for a weakly interacting
Bose gas of polaritons at $n(0)a_{2D}^{2} \ll 1$, where $n(0)$ is the density
of polaritons at the center of the trap. In this figure it is clearly evident
that a Bose-Einstein condensate exists in the polariton system up to $T>20$~K,
i.e.\ at temperatures much higher than the temperature of the
normal-to-superfluid transition in $^4$He, $T_{\lambda} = 2.17$ K.

\subsection{Trapped polaritons in graphene
in a microcavity in high magnetic fields}

Recent advances in the fabrication  of carbon-based materials have enabled the
production of graphene, that is, a two-dimensional honeycomb lattice of carbon
atoms forming the basic planar structure \cite{Novoselov1,Zhang1,guinea}.
Graphene is a semi-metal that is characterized by an unusual electronic
structure. In particular, the electronic excitations with energy close to the
Fermi energy (the last occupied energy level) have a linear dispersion with
virtually zero effective mass. Therefore, these excitations in graphene
represent a condensed-matter analogue of Dirac fermions (like neutrinos) in
standard quantum field theories \cite{Novoselov2}. Because of this, and due to
high potential for the use in electronic devices, graphene has stimulated a
considerable amount of theoretical and experimental study
\cite{Novoselov2,Zhang2,Zhang3}. In this Section, we briefly describe the
properties of trapped polaritons is a single graphene layer (GL) in an optical
microcavity in a high magnetic field (i.e.\ magnetopolaritons: see below). In
particular, we show that this system can undergo a superfluid transition at
finite temperature ~\cite{Berman_magnetopolaritons}.

It is known \cite{Landau:65} that charged particles can readily be trapped in a
potential well in an external magnetic field directed perpendicularly to the
layer. Because of the quantization of particle motion in a magnetic field, the
energy of the charged particles can only take discrete values, called Landau
levels, at $E_n = (n + 1/2) \hbar \omega_B - 2 \mu_B s_z B$, where $B$ is the
magnetic field strength, $n$ is the level number, and $\omega_B =  |e| B / m c$
is the cyclotron frequency \cite{Landau:65}, $m$ is the mass of the particle,
$\mu_B$ is the Bohr magneton, and $s_z$ is a projection of the electronic spin
on the direction of the magnetic field. In a magnetic field, multiple electrons
can occupy the same Landau level. The degeneracy of an individual Landau level
i.e.\ the maximum number of electrons on it, is proportional to the magnetic
field strength, $\nu = |e| B S / 2 \pi \hbar c$ with $S$ being the layer area
\cite{Landau:65,Landau:statphys}. In full analogy with consideration given in
Sec. \ref{sec:bec} above, the electrons and the holes in a graphene layer in
magnetic field are coupled into excitons, which are referred to as
\textit{magnetoexcitons} ~\cite{Berman_magnetopolaritons}.

In undoped graphene, in a perpendicular magnetic field (without an external
electric field) and in the ground state, half of the 0th Landau level, i.e.\
$\nu/2$ states, is occupied by electrons, whereas all Landau levels at $n>0$
are empty. We label the Landau levels for electrons by numbers $n\geq 0$, and
the Landau levels for the holes that are the positive charge carriers, by
$n\leq 0$. Thus, the chemical potential of the system is positioned at the
level $n=0$.

In the presence of an electric field, which is  produced in experiments by
applying the ``gate'' voltage to the system, the chemical potential of the
magnetoexcitons can be changed in either of two ways: (a) an increase in the
chemical potential in such a way that the resulting chemical potential is
positioned between the 0th and 1st Landau levels;  or (b) a decrease in the
chemical potential so that it is positioned between the first negative ($-1$)
and 0th Landau levels \cite{Gusynin3}. In case (a) the magnetoexcitons are
formed from electrons in the 1st Landau level and holes in the 0th Landau
level, whereas in case (b) the magnetoexcitons are formed from electrons in the
0th Landau level and holes in the Landau level at $n = -1$. (Note that, by
applying an appropriate gate voltage we can also use any other neighboring
Landau levels at $n$ and $n+1$.) In both cases, all Landau levels below the
chemical potential are completely occupied and all Landau levels above the
chemical potential are completely empty.

For a relatively high dielectric constant $\epsilon$ of the microcavity, the
magnetoexciton energy can be calculated by applying perturbation theory with
the small parameter being the strength of the Coulomb electron-hole attraction
~\cite{Berman_magnetopolaritons}. The computation  is similar to that
~\cite{Lerner} made for electrons and holes in a 2D quantum well in high
magnetic fields. This approach is valid if the Coulomb electron-hole
interaction energy in the single graphene layer, $e^{2}/(\epsilon r_B)$ is
smaller than the energy difference between the Landau levels in graphene,  $
\hbar v_{F}/r_b$, where $r_B = (\hbar c/|e|B)^{1/2}$ denotes the magnetic
length of the magnetoexcitons in the magnetic field $B$, $v_{F}$ is the Fermi
velocity, that is, the velocity of electrons of energy equal to the Fermi
energy. This requirement is satisfied, for example, for graphene embedded in a
GaAs microcavity, for which $\epsilon \approx 13$.  (Note, that we restrict
ourselves to consideration of high magnetic fields.)

The effective Hamiltonian of the  magnetopolaritons trapped
by the potential $V(r)$ in the graphene layer embedded in a microcavity is
\begin{eqnarray}
\label{Ham_eff} \hat H_{\rm eff}  =
\sum_{\mathbf{P}}\left(\frac{P^{2}}{2M_{\rm eff}^{\rm (mp)}(B)} + \frac{1}{2}
V(r) \right)\hat{p}_{\mathbf{P}}^{\dagger}\hat{p}_{\mathbf{P}} \ ,
\end{eqnarray}
where $\hat{p}_{\mathbf{P}}^{\dagger}$ and $\hat{p}_{\mathbf{P}}$ are the lower
polariton creation and annihilation  operators (cf.\ Sec. \ref{sec:bec}). We
see that the magnetic field strength only enters into the Hamiltonian
(\ref{Ham_eff}) through the effective mass of the magnetopolaritons, which is
\begin{eqnarray}
\label{Meff_pol} M_{\rm eff}^{\rm (mp)}(B) = 2   \left(m_{B}^{-1} + \frac{c
\lambda_C (B)}{n\hbar\pi}\right)^{-1} \ .
\end{eqnarray}
Here, $\lambda_C$ is the length of the cavity, and  $m_{B} = 2^{7/2} \epsilon
\hbar^2 / \pi ^{1/2} e^2 r_B$ is the magnetoexciton mass in graphene
\cite{Berman_magnetopolaritons}. We suppose that the magnetic field is chosen
so that the magnetoexciton and photon modes are in mutual resonance. Therefore,
the length decreases with increasing magnetic field as $\lambda_C \propto
B^{-1/2}$.

From Eq.~(\ref{Meff}) it follows that the effective magnetopolariton mass
$M_{\rm eff}^{\rm (mp)}$ increases with the magnetic field as $B^{1/2}$. The
Hamiltonian (\ref{Ham_eff}) is similar to that for the pure polariton system
(\ref{Ham_eff_real}). Meanwhile, it can be shown that in a high magnetic field
the interaction between magnetopolaritons is negligibly small
~\cite{Berman_magnetopolaritons} and, because of that, that the fourth order
term on the r.h.s. of Eq.\  (\ref{Ham_eff}) can be omitted. Thus the
magnetopolaritons in a quantum well in a graphene layer form a virtually ideal
quantum gas.

As explained in Sec. \ref{sec:bec}, two-dimensional Bose particles in a
trapping potential undergo Bose-Einstein condensation at  some critical
temperature $T=T_{c}^{(0)}$. Computations similar to those made in Sec.\
\ref{sec:bec} give the following expression for the condensate fraction of
magnetopolaritons at $T < T_{c}^{(0)}$ \cite{Berman_magnetopolaritons},
\begin{equation}
 \label{n_con}
N_{0}(T,B) = N - \frac{\pi  \left(g_{s}^{(e)}g_{v}^{(e)} +
 g_{s}^{(h)}g_{v}^{(h)}\right)M_{\rm
eff}(B)}{3\hbar^{2}\gamma}(k_{B}T)^{2}.
\end{equation}
In Eq. (\ref{n_con}) $N$ is the total number of magnetopolaritons, and
$g_{s}^{(e),(h)}$ and $g_{v}^{(e),(h)}$ are the spin and momentum degeneracy
factors  for electrons and holes in graphene, respectively. At $T=T_{c}^{(0)}$
the condensate fraction vanishes. Setting $N_{0}=0$ in Eq. (\ref{n_con}), we
estimate the critical temperature $T_{c}^{(0)}$ for the ideal gas of
magnetopolaritons as
\begin{eqnarray}
\label{t_c}
T_{c}^{(0)} (B) = \frac{1}{k_{B}}\left(\frac{3\hbar^{2}\gamma N}{8 \pi
M_{\rm eff}(B)} \right)^{1/2} \ .
\end{eqnarray}
According to Eq.~(\ref{t_c}), the  critical temperature $T_{c}^{(0)}$ decreases
with magnetic field  as $B^{-1/4}$ and increases with spring constant as
$\gamma^{1/2}$. Fig.~\ref{N0} (right) demonstrates the dependence of the
normalized critical temperature for the magnetopolariton condensate on external
magnetic field. In the computations, we used $g_{s}^{(e)} = g_{v}^{(e)} =
g_{s}^{(h)} = g_{v}^{(h)} = 2$.

It is important to note that taking account of higher order terms in the
Hamiltonian (\ref{Ham_eff}) (which are of the relative order of $1/\epsilon$)
results in magnetopolariton-magnetopolariton interactions and hence
superfluidity in the magnetopolariton system.

\subsection{Drag effects in the system of electrons and microcavity
polaritons}

Because the exciton consists of a coupled electron and hole, and hence has zero
net charge, it can not be moved directly by application of an external electric
field. However, an electrical current in a neighboring, spatially separated,
quantum well can entrain excitons and therefore induce their
motion~\cite{BKL_drag_PRB}. In this subsection, we consider such drag effects
for polaritons in a quantum well in a microcavity and propose possible
experiment on the generation of collective polariton flow. Moreover, because
the excitons are entangled with the cavity photons,  polariton motion causes
changes in the distribution of light emitted from the microcavity. In effect,
we can direct light by application of voltage to the quantum well
~\cite{BKL_drag_PLA}.

\begin{figure}[t] \centering
\includegraphics[width=70mm]{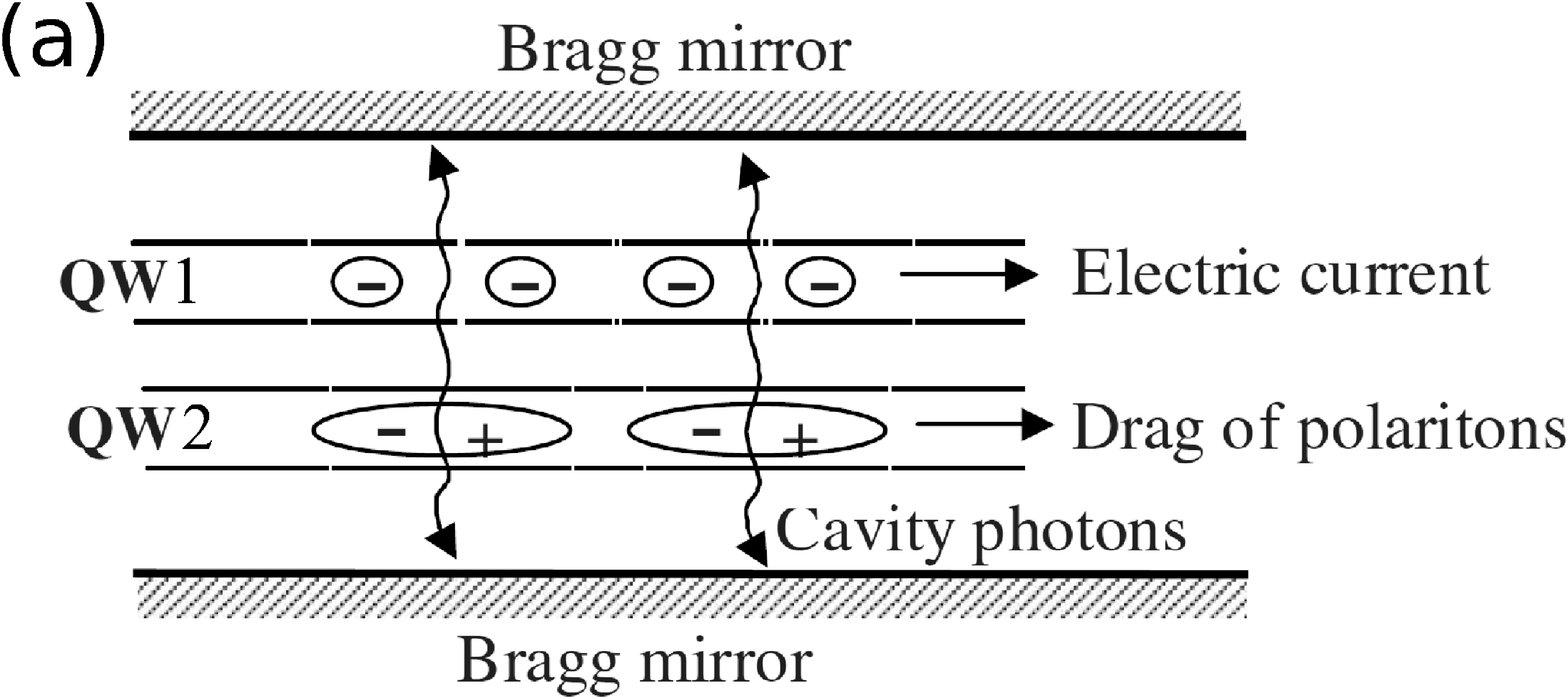}
\includegraphics[width=70mm]{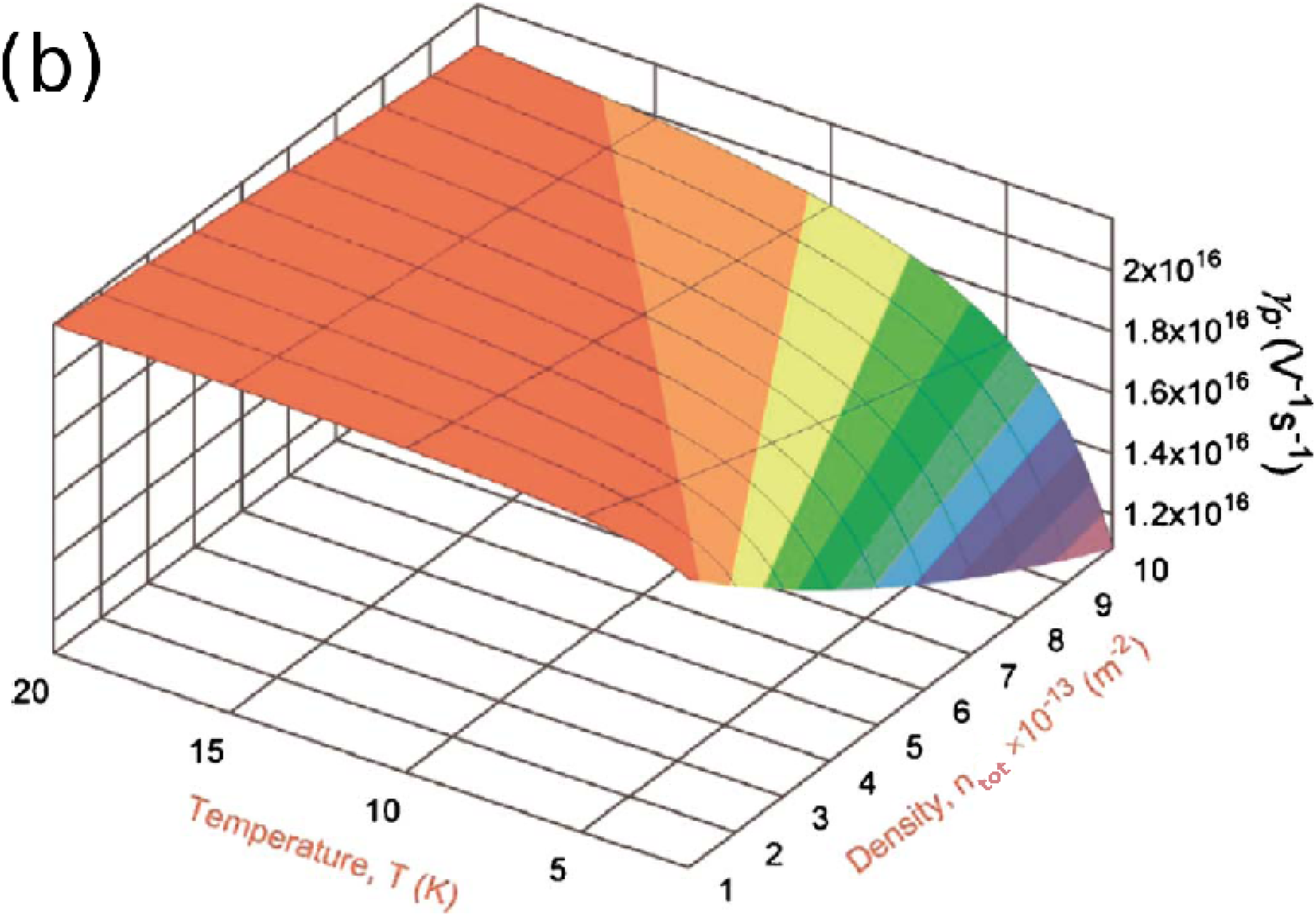}
\caption{(a) The schematic diagram for the drag effects in quantum wells
embedded in an optical microcavity. At low temperatures the polaritons
in the lower quantum well QW2 are dragged by the electron current in the upper quantum well
QW1. (b) The drag coefficient $\gamma_p$ in $V^{-1} s^{-1}$ in the
system of superfluid microcavity polaritons and electrons as a function
of temperature $T$ in K and total polariton density $n_{tot}$ in $m^{-2}$. The
interwell separation is $D=20$ nm. We used the parameters for the
GaAs/GaAsAl quantum wells: $m_e=0.07m_0$, $m_h=0.15m_0$, $M
=0.24m_0$, and $\epsilon = 13$.  From Ref.  \cite{BKL_drag_PRB}} \label{dr}
\end{figure}

In this setup, we consider two neighboring quantum wells embedded in an optical
microcavity as shown in Fig.\ \ref{dr}(a). The ``upper'' quantum well (QW1) is
occupied by 2D electron gas, and the ``lower'' quantum well (QW2) is occupied
by the excitons created by laser pumping. It is important to note that the
quantum wells are not identical: QW1 are doped so that there are excessive free
negative charges, whereas in QW2 the number of negative and positive charges
are equal to each other. We will consider the case of relatively low
temperature where the occupation of the upper polariton branch is exponentially
small, that is $k_{B}T \ll \hbar\Omega_{R}$ where $\Omega_{R}$ is Rabi
splitting, which is equal to half the transition frequency from the upper to
lower polariton branches at $k_{||}=0$ (see Fig.\ \ref{cav}({\it b})).  At such
temperatures the polaritons in QW2 form a superfluid \cite{BKL_drag_PRB}. In
the computations, we omit small effects due to the nonresonant interaction of
photons in the microcavity with electrons in the upper quantum well QW1.

By applying an electric voltage, an electron current is induced in QW1 (its
direction is shown in Fig.\ \ref{dr} by the arrow). This current drags the
normal component of excitons in the  neighboring QW2 due to polarization effects
~\cite{BKL_drag_PLA,BKL_drag_PRB}.

This entrainment effect can be characterized  by the drag coefficient
of the polaritons, $\gamma_p$, introduced as follows \cite{BKL_drag_PRB}
\begin{equation}
\bm{i}_p = -D_p \nabla n_p + \gamma_p \bm{E}, \label{dragcoeff}
\end{equation}
where $\bm{i}_p$ is the mass current of the polaritons, $D_p$ the polariton
diffusion coefficient in the QW2 plane, $n_p$ is the two-dimensional  density
of thermally activated quasiparticles, which form the normal component of the
polariton superfluid, and $\bm{E}$ is the  lateral electric field in the upper
quantum well QW1. Note that the drag effect is fully reversible: the
polaritonic current formed in the quantum well QW2 will generate a non-zero
current of electrons in QW1  \cite{BKL_drag_PRB}.

The dependence of the drag coefficient $\gamma_p$ on the total density
$n_{tot}$ of the temperature-activated quasiparticles in the polariton
superfluid, and on the temperature $T$, is demonstrated in Fig.\ \ref{dr}(b)
\cite{BKL_drag_PRB}. It follows from the figure that the drag coefficient grows
with increasing temperature $T$ and slowly decreases with increasing density
$n_{tot}$. It can be demonstrated by detailed computations \cite{BKL_drag_PRB}
that the dependence of the $\gamma_p$ coefficient on the both parameters is
defined mainly by variations in the density of the normal component of the
superfuid. In particular,  $\gamma_p$ tends to zero in the limit of very low
temperatures, at which the normal component vanishes. We have to note that $T$
here represents an average temperature of the system; the exact, local
temperature varies in the direction of the electric field due to changes in the
concentration of the quasiparticles  (i.e.\ due to non-zero gradient of the
normal density in the polariton superfluid.)

Generation of the lateral polaritonic current due to the drag effect causes
crucial changes in the photoluminiscence spectrum, that is, in the angular
distribution of the photons leaving the microcavity. We recall that the mirrors
in the setup shown in Fig. \ref{cav}(a) are semi-transparent. Fig. 8
demonstrates the angular dependence of the intensity $F(\theta)$ of photons
emitted by the excitonic superfluid. We defined  $\theta$ as an angle between
the $\bm{k}$ vector of the  emitted photon and the unit vector normal to the
Bragg mirrors. As illustrated in Fig. 8, a sharp peak in the photon intensity
$F(\theta)$ caused by photoluminiscence from the superfluid component is
positioned at $\theta=0$ in the cases of both zero (Fig. 8(a)) and  finite
(Fig. 8(b)) polariton current $\bm{i}_p$. Nevertheless, the broad background
distribution in $F(\theta)$, which is generated by photoluminiscence of the
normal polaritonic component, is shifted for finite polariton current compared
to the case of zero polariton current. In other words, we can control the angle
at which the light is emitted from the microcavity by generating an electronic
current in the quantum well QW1.

This effect opens up an opportunity for designing thresholdless tunable
micro-lasers, which may be of importance for applications in modern quantum
electronics \cite{BKL_drag_PLA,BKL_drag_PRB}. Besides, the control of photons
and/or excitons by the exciton-electron drag can be used for studies of the
properties of the polariton and exciton system and, in particular, of
superfluidity in the system.


\begin{figure}[t]
\hspace{1.cm}\begin{minipage}{80.mm}
\includegraphics[width=80.mm]{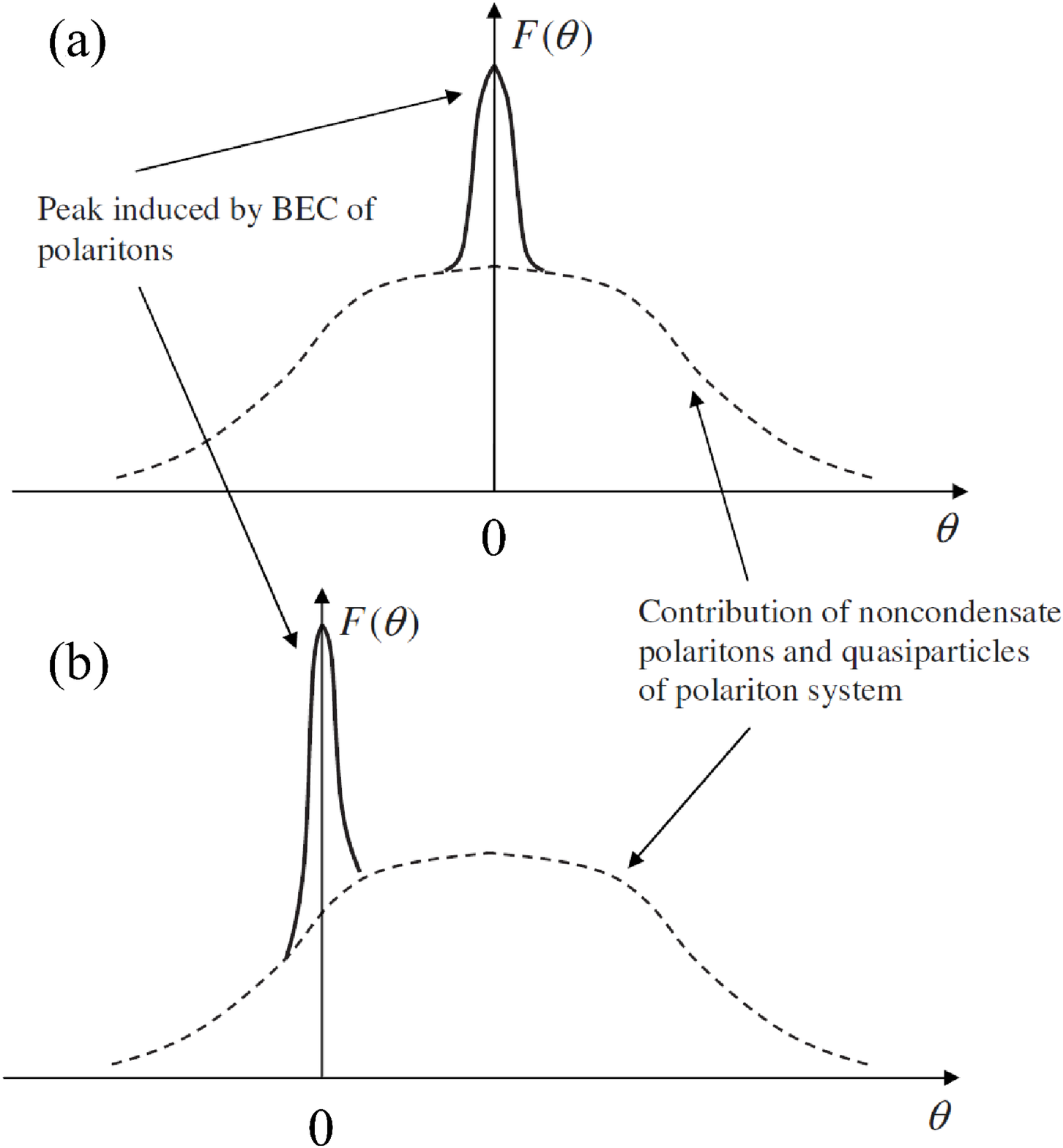}
\end{minipage}\hspace{1.cm}
\begin{minipage}[c]{0.3\linewidth}
\caption{Proposed experiment: the manifestation of polariton drag
effect through change of the angular distribution of the photons
escaping the optical microcavity. (a) The angular distribution of the
photons escaping the optical microcavity without drag. (b) The
angular distribution of the photons escaping the optical microcavity
in the presence of the drag effect. $F(\theta)$ is the
light intensity emitted from the microcavity at an angle $\theta$
with respect to the direction normal to the Bragg mirrors. From Ref.
\cite{BKL_drag_PRB}}
\end{minipage}\label{exper}
\end{figure}

\subsection{Discussion}
We presented above the current status of theoretical research of BEC and
superfluidity of trapped quantum well polaritons in a microcavity. Besides, we
report BEC of trapped magnetoexciton polaritons in GL and QW  embedded in an
optical microcavity in high magnetic field. In the both cases the polaritons
are considered in a harmonic potential trap. The effective Hamiltonian of QW
and GL polaritons in a microcavity in a high magnetic field, and the BEC
temperature as functions of magnetic field, are obtained. It is shown that the
effective magnetic mass of the magnetoexciton polariton depends on magnetic
field.

We have also discussed drag effects in the system of spatially separated
electrons and excitons in coupled quantum wells embedded in an optical
microcavity were discussed, and the corresponding experiments.

We considered how, for low temperatures, the Hamiltonian of the 2D exciton
polaritons in a slowly varying external parabolic potential acts on the exciton
energy and brings it into resonance with a cavity photon mode, corresponding
directly to the case of a weakly interacting Bose gas with an effective mass
and effective pairwise interaction in a harmonic potential trap. The condensate
fraction is a decreasing function of temperature, as expected, and an
increasing function of the curvature of the parabolic potential. Mixing with
the photon states leads to a smaller lifetime for high-energy states, that is,
an evaporative cooling effect, but it does not fundamentally prevent
condensation. Since harmonic potential traps are now possible for microcavity
polaritons, it should be possible to compare these calculations with
experimental results for the critical density and spatial profile of the
polariton condensate.

Since microcavity polaritons form a BEC, light emitted from the condensate due
to electron-hole pair recombination is nearly coherent. This coherence of the
emitted light means that polariton BEC in a microcavity can be used in
technology as polariton laser, since photon energy is pumped in incoherently,
putting electrons into the excited states, and coherent light is
emitted~\cite{Lit,Snoke_text}. Moreover, because the emitted light frequency
for the graphene-based polariton laser depends on the external magnetic field
strength, magnetopolariton BEC in graphene in microcavity can be used for the
design of a polariton laser tunable by magnetic field.

\section{Conclusions}\label{sec:Conclusions}
In conclusion, we have presented an introductory review of the non-equilibrium
dynamics of quantum systems, a fast developing field. We devise an approach to
the description of the out-of-equilibrium quantum liquid based on a Hamiltonian
formalism of superfluid hydrodynamics. Through analysis of the quasiclassical
kinetic equations for the correlation functions for the first and second sound
wave modes we show that, in the non-equilibrium state i.e.\ where there is a
constant energy input into the superfluid, the time evolution of the system is
affected by nonlinearity. As a result of nonlinear wave interaction, the direct
and inverse cascades corresponding to constant fluxes of the integrals of
motion (the energy and the wave action), are established in the bulk
superfluid. The stationary solutions of the kinetic equations that describe
this state are similar to those that appear in classical nonlinear wave
dynamics. However, in contrast to the classical picture, both the direct and
inverse cascades appear as a result of three-wave interactions between sound
modes. We demonstrate that analytic and numerical approaches describe well the
results of experiments with nonlinear waves in superfluid $^4$He. We also
analyze the conditions under which Bose-Einstein condensate forms in the system
of interacting exciton polaritons in optical microcavities in semicondictors,
and in a graphene layer. In this work, the occupation numbers of
out-of-condensate quantum states are not large and the effect of quantum
fluctuations is important; we therefore use an operator formulation of the
Hamiltonian approach. We show that because of small effective mass of the
polaritons, the Bose-Einstein condensate in this system can form at a
temperature much higher than the superfluid transition temperature in bulk
liquid helium. We also studied the effect of an external magnetic field on the
BEC condensation. The results obtained in this part are directly applicable to
the systems and materials that are potentially useful for the design of novel
microelectronic and of micro-lasing devices.

\section*{Acknowledgments}
G. Kolmakov is grateful to the Center for Theoretical Physics at the New York
City College of Technology CUNY for hospitality. The authors are grateful to P.
V. E. McClintock, L. P. Mezhov-Deglin, V. E. Zakharov, A. A. Levchenko, A. N.
Ganshin and V. B. Efimov for valuable discussions. The investigations were
supported in part by the National Science Foundation U.S.A. through Teradgid
grant and by the Engineering and Physical Sciences Research Council (U.K.).


\end{document}